\documentclass[5p, authoryear]{elsarticle}
\usepackage[utf8]{inputenc} 
\usepackage{graphicx}
\usepackage{rotating}	
\usepackage{setspace}
\usepackage{natbib}
\usepackage{aas_macros}
\usepackage{hyperref}
\usepackage{amsmath}
\usepackage{amssymb}
\usepackage{todonotes}
\usepackage{longtable}
\usepackage{lineno}
\hypersetup{colorlinks=true,linkcolor=blue,citecolor=blue,urlcolor=blue} 

\DeclareGraphicsExtensions{.png, .pdf, .jpeg, .eps} 
\newcommand{\methane}{$\mathrm{CH_4}$}
\newcommand{\methyla}{$\mathrm{C_3H_4}$}
\newcommand{\cyano}{$\mathrm{C_2N_2}$}
\newcommand{\diacety}{$\mathrm{C_4H_2}$}

\title{Seasonal evolution of temperatures in Titan's lower stratosphere}
\author[bristol]{M. Sylvestre\corref{cor1}}
\ead{melody.sylvestre@bristol.ac.uk}
\author[bristol]{N. A. Teanby}
\author[lmd]{J. Vatant d'Ollone}
\author[meudon]{S. Vinatier}
\author[meudon]{B. Bézard}
\author[lmd]{S. Lebonnois}
\author[oxford]{P.~G.~J.~Irwin}

\address[bristol]{School of Earth Sciences, University of Bristol, Wills Memorial Building, Queens Road,\\
	Bristol BS8 1 RJ, UK }
\address[lmd]{Laboratoire de Météorologie Dynamique (LMD/IPSL), Sorbonne  Université, ENS, PSL Research University, Ecole Polytechnique, Université Paris Saclay, CNRS, 4 Place Jussieu, F 75252 Paris Cedex 05, France}
\address[meudon]{LESIA, Observatoire de Paris, Université PSL, CNRS, Sorbonne Université, Univ. Paris Diderot, Sorbonne Paris Cité, 5 place Jules Janssen, 92195 Meudon, France}
\address[oxford]{Atmospheric, Oceanic, \& Planetary Physics, Department of Physics, University of Oxford, Clarendon Laboratory, Parks Road, Oxford OX1 3PU, UK}

\cortext[cor1]{Corresponding author}
\begin{document}

\begin{abstract}
	The Cassini mission offered us the opportunity to monitor the seasonal evolution of Titan's atmosphere from 2004 to 2017, i.e. half a Titan year. The lower part of the stratosphere (pressures greater than 10~mbar) is a region of particular interest as there are few available temperature measurements, and because its thermal response to the seasonal and meridional insolation variations undergone by Titan remain poorly known. In this study, we measure temperatures in Titan's lower stratosphere between 6~mbar and 25~mbar using Cassini/CIRS spectra covering the whole duration of the mission (from 2004 to 2017) and the whole latitude range. We can thus characterize the meridional distribution of temperatures in Titan's lower stratosphere, and how it evolves from northern winter (2004) to summer solstice (2017). Our measurements show that Titan's lower stratosphere undergoes significant seasonal changes, especially at the South pole, where temperature decreases by 19~K at 15~mbar in 4 years.\\

\end{abstract}
\maketitle

\section{Introduction}

Titan has a dense atmosphere, composed of $\mathrm{N_2}$ and \methane, and many trace gases such as  hydrocarbons (e.g. $\mathrm{C_2H_6}$, $\mathrm{C_2H_2}$) and nitriles (e.g. $\mathrm{HCN}$, $\mathrm{HC_3N}$) produced by its rich photochemistry.  Like Earth, Titan has a  stratosphere, located between 50~km ($\sim 100$~mbar) and 400~km  ($\sim 0.01$~mbar), characterized by the increase of its temperature with altitude because of the absorption of incoming sunlight by methane and hazes. Titan's atmosphere undergoes strong variations of insolation, due to its obliquity ($26.7^{\circ}$) and to the eccentricity of Saturn's orbit around the Sun (0.0565). \\

The Cassini spacecraft monitored Titan's atmosphere during 13 years (from 2004 to 2017), from northern winter to summer solstice. Its data are a unique opportunity to study the seasonal evolution of its stratosphere, especially with mid-IR observations from Cassini/CIRS (Composite InfraRed Spectrometer, \citet{Flasar2004}). They showed that at pressures lower than 5~mbar, the stratosphere exhibits strong seasonal variations of temperature and composition related to changes in atmospheric dynamics and radiative processes. For instance, during northern winter (2004-2008), high northern latitudes were enriched in photochemical products such as HCN or \diacety, while there was a "hot spot" in the upper stratosphere and mesosphere (0.1 - 0.001~mbar, \citet{Achterberg2008, Coustenis2007, Teanby2007b,Vinatier2007}). These observations were interpreted as evidence of subsidence above the North pole during winter, which is a part of the pole-to-pole atmospheric circulation cell predicted for solstices by Titan GCMs (Global Climate Models, \citet{Lora2015,Lebonnois2012a,Newman2011}). These models also predict that the circulation pattern should reverse  around equinoxes, via a transitional state with two equator-to-pole cells. These changes began to affect the South pole in 2010, when measurements showed that pressures inferior to 0.03~mbar exhibited an enrichment in gases such as HCN or $\mathrm{C_2H_2}$, which propagated downward during autumn, consistent with the apparition of a new circulation cell with subsidence above the South pole \citep{Teanby2017,Vinatier2015}.\\

 Some uncertainties remain about the seasonal evolution of the lower part of the stratosphere, i.e. at pressures from 5~mbar (120~km) to 100~mbar (tropopause, 50~km). Different estimates of radiative timescales have been calculated for this region. In \citet{Strobel2010}, the radiative timescales in this region vary from 0.2 Titan years at 5~mbar to 2.5 Titan years at 100~mbar. This means that the lower stratosphere should be the transition zone from parts of the atmosphere which are sensitive to  seasonal insolation variations, to parts of the atmosphere which are not.  In contrast, in the radiative-dynamical model of \citet{Bezard2018}, radiative timescales are between 0.02 Titan year at 5~mbar and 0.26 Titan year at 100~mbar, implying that this whole region should exhibit a response to the seasonal cycle.\\

From northern winter to equinox, CIRS mid-IR observations showed that temperature variations were lower than 5~K between 5~mbar and 10~mbar \citep{Bampasidis2012,Achterberg2011}.  Temporal variations intensified after spring equinox, as \citet{Coustenis2016} measured a cooling by 16~K and an increase in gases abundances at $70^{\circ}$S from 2010 to 2014, at 10~mbar, associated with the autumn subsidence above the South pole. \citet{Sylvestre2018} showed that this subsidence affects pressure levels as low as 15~mbar as they measured  strong enrichments in \cyano, \methyla, and \diacety~at high southern latitudes from 2012 to 2016 with  CIRS far-IR observations. However, we have little information on temperatures and their seasonal evolution for pressures greater than 10~mbar. Temperatures from the surface to 0.1~mbar can be measured by Cassini radio-occultations, but the published profiles were measured mainly in 2006 and 2007 \citep{SchinderFlasarMaroufEtAl2011,Schinder2012}, so they provide little information on seasonal variations of temperature. \\ 
 
In this study, we analyse all the available far-IR Cassini/CIRS observations to probe temperatures from 6~mbar to 25~mbar, and measure the seasonal  variations of  lower stratospheric temperatures. As these data were acquired throughout the Cassini mission from 2004 to 2017, and cover the whole latitude range, they provide a unique overview of the thermal evolution of the lower stratosphere from northern winter to summer solstice, and a better understanding of the radiative and dynamical processes at play in this part of Titan's atmosphere.\\

\section{Data analysis}	

	\subsection{Observations}
	 We measure lower stratospheric temperatures using Cassini/CIRS \citep{Flasar2004} spectra. CIRS is a thermal infrared spectrometer with three focal planes operating in three different spectral domains: 10 - 600$~\mathrm{cm^{-1}}$ (17 - 1000$~\mathrm{\mu m}$) for FP1,  600 - 1100$~\mathrm{cm^{-1}}$ (9 - 17 $~\mathrm{\mu m}$) for FP3, and 1100 - 1400$~\mathrm{cm^{-1}}$ (7 - 9$~\mathrm{\mu m}$) for FP4. FP1 has a single circular detector with an angular field of view of 3.9~mrad, which has an approximately Gaussian spatial response with a FWHM of 2.5 mrad. FP3 and FP4 are each composed of a linear array of ten detectors. Each of these detectors has an angular field of view of 0.273~mrad. \\
	
\begin{figure}[!h]
	\includegraphics[width=1\columnwidth]{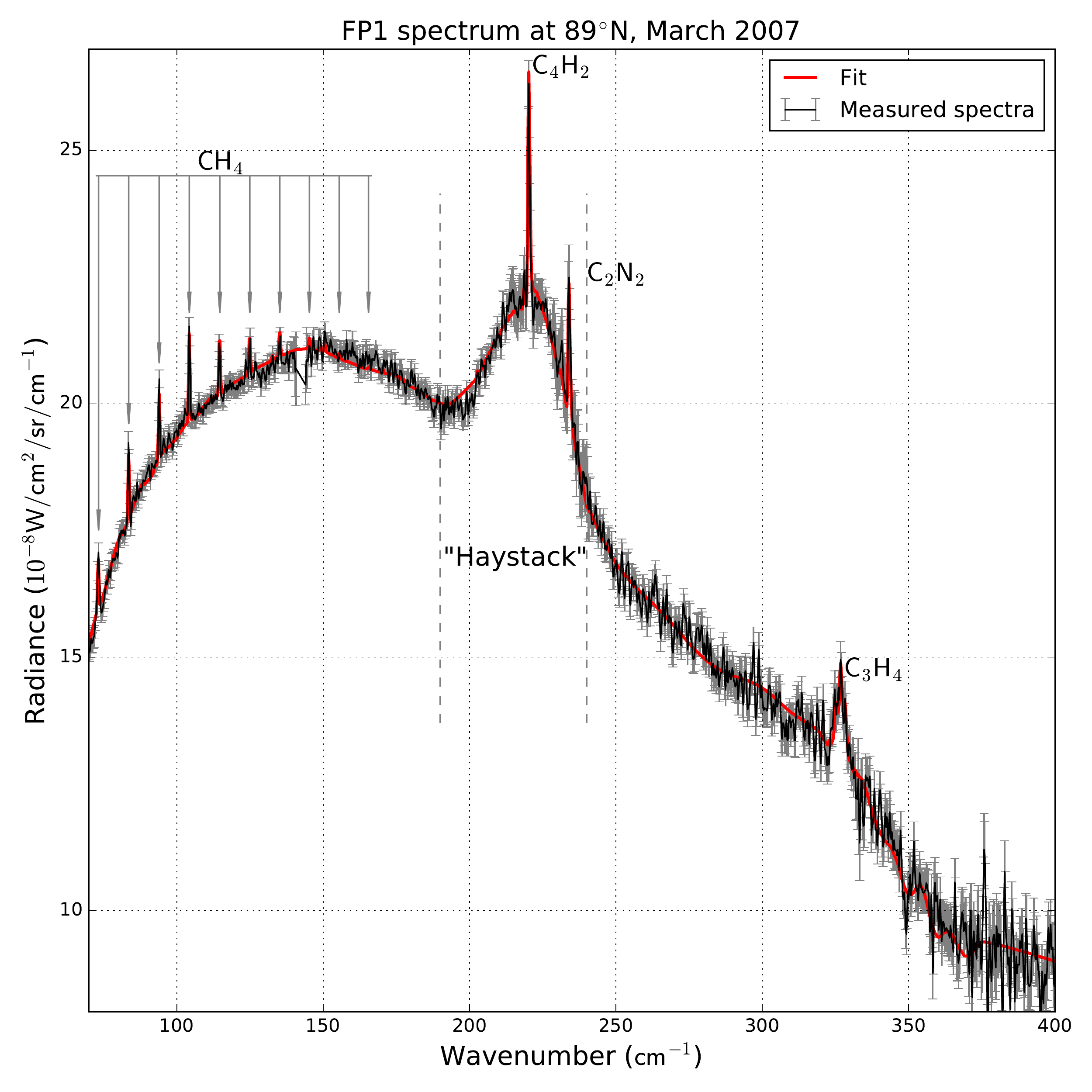}
	\caption{Example of average spectrum measured with the FP1 detector of Cassini/CIRS (in black) and its fit by NEMESIS (in red). The measured spectrum was obtained after averaging 106 spectra observed at $89^{\circ}$N in March 2007. The rotational lines of \methane~are used to retrieve stratospheric temperature. The "haystack" feature is visible only at high latitudes during autumn and winter. }
	\label{fig_spec}
\end{figure}

	 In this study,  we use FP1 far-IR observations, where nadir spectra are measured at a resolution of 0.5$~\mathrm{cm^{-1}}$,  in  "sit-and-stare" geometry (i.e the FP1 detector probes the same latitude and longitude during the whole duration of the acquisition). In this type of observation, the average spatial field of view is 20$^\circ$ in latitude. An acquisition lasts between 1h30 and 4h30, allowing the recording of 100 to 330 spectra. The spectra from the same acquisition are averaged together, which increases the S/N by a factor $\sqrt{N}$ (where N is the number of spectra).  As a result, we obtain an average spectrum where the rotational lines  of \methane~(between 70$~\mathrm{cm^{-1}}$ and 170$~\mathrm{cm^{-1}}$) are resolved and can be used to retrieve Titan's lower stratospheric temperature. An example averaged spectrum is shown in Fig. \ref{fig_spec}.\\
	 
	 We analysed all the available observations with the characteristics mentioned above. As shown in table \ref{table_obs}, this type of nadir far-IR observation has been performed throughout the Cassini mission (from 2004 to 2017), at all latitudes.  Hence, the analysis of this dataset enables us to get an overview of Titan's lower stratosphere and its seasonal evolution.\\ 

	\subsection{Retrieval method}
		 We follow the same method as \citet{Sylvestre2018}. We use the portion of the spectrum between  70~$\mathrm{cm^{-1}}$ and 400~$\mathrm{cm^{-1}}$, where the main spectral features are: the ten rotational lines of \methane~(between 70$~\mathrm{cm^{-1}}$ and 170$~\mathrm{cm^{-1}}$), the \diacety~band at $220~\mathrm{cm^{-1}}$,  the \cyano~band at  $234~\mathrm{cm^{-1}}$, and the \methyla~band at  $327~\mathrm{cm^{-1}}$ (see Fig. \ref{fig_spec}). The continuum emission comes from the collisions between the three main components of Titan's atmosphere (N$_2$, \methane, and H$_2$), and from the spectral contributions of the hazes. \\
		
		We retrieve the temperature profile using the constrained non-linear inversion code NEMESIS \citep{Irwin2008}. We define a reference atmosphere, which takes into account the abundances of the main constituents of Titan's atmosphere measured by Cassini/CIRS \citep{Coustenis2016,Nixon2012,Cottini2012,Teanby2009}, Cassini/VIMS \citep{Maltagliati2015}, ALMA \citep{Molter2016} and Huygens/GCMS\citep{Niemann2010}. We also consider the haze distribution and properties measured in previous studies with Cassini/CIRS \citep{deKok2007,deKok2010b,Vinatier2012}, and Huygens/GCMS \citep{Tomasko2008b}. We consider four types of hazes, following \citet{deKok2007}:  hazes 0 ($70~\mathrm{cm^{-1}}$ to $400~\mathrm{cm^{-1}}$), A (centred at $140~\mathrm{cm^{-1}}$), B (centred at $220~\mathrm{cm^{-1}}$) and C (centred at  $190~\mathrm{cm^{-1}}$).  For the spectra measured at high northern and southern latitudes during autumn and winter, we add an offset from 1 to $3~\mathrm{cm^{-1}}$  to the nominal haze B cross-sections between 190~$\mathrm{cm^{-1}}$ and 240~$\mathrm{cm^{-1}}$, as in \citet{Sylvestre2018}.  This modification improves the fit of the continuum in the "haystack" which is a strong emission feature between 190~$\mathrm{cm^{-1}}$ and 240~$\mathrm{cm^{-1}}$ (see Fig. \ref{fig_spec}) seen at high latitudes during autumn and winter (e.g. in  \citet{Coustenis1999, deKok2007, Anderson2012, Jennings2012, Jennings2015}). The variation of the offset allows us to take into account the evolution of the shape of this feature throughout autumn and winter. The composition of our reference atmosphere and the spectroscopic parameters adopted for its constituents are fully detailed in \citet{Sylvestre2018}.\\
			
		We retrieve the temperature profile and scale factors applied to the \textit{a priori} profiles of \cyano, \diacety, \methyla, and hazes 0, A, B and C, from the spectra using the constrained non-linear inversion code NEMESIS \citep{Irwin2008}. This code generates synthetic spectra from the reference atmosphere. At each iteration, the difference between the synthetic and the measured spectra is used to modify the profile of the retrieved variables, and minimise a cost function, in order to find the best fit for the measured spectrum. \\
		
\begin{figure}[!h]
		\includegraphics[width=1\columnwidth]{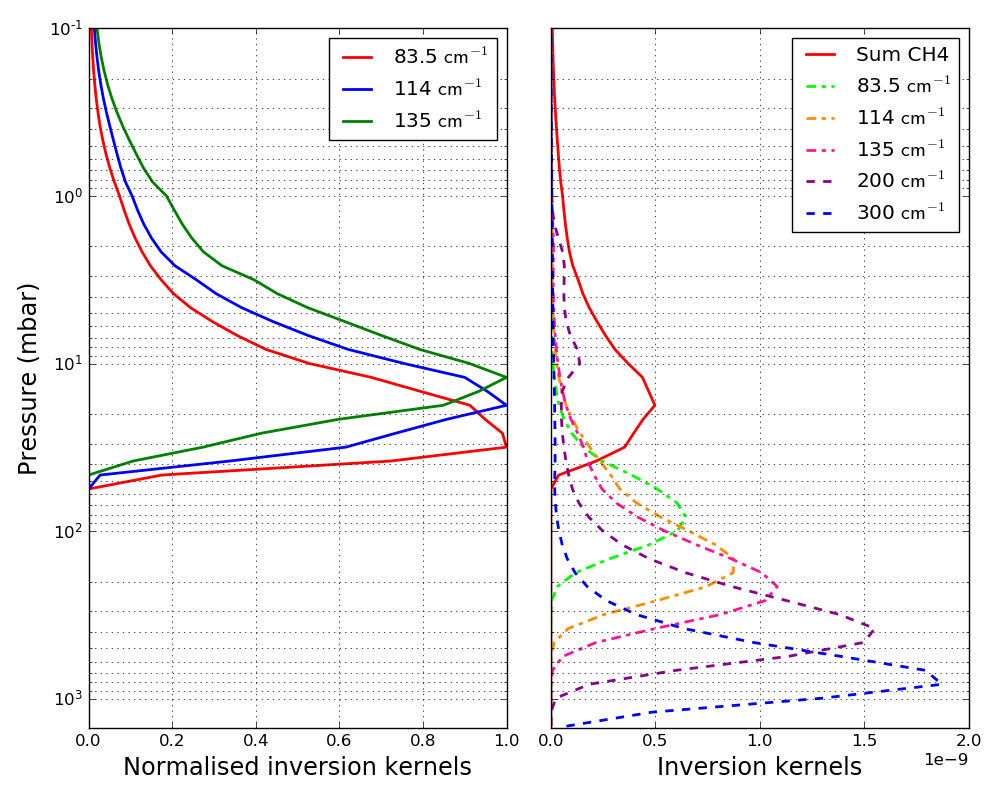}
		\caption{Sensitivity of temperature measurements at $72^{\circ}N$ in April 2007. \textit{Left panel}: Normalised inversion kernels $K^{CH_4}_{ij}$ in three of the \methane~rotational lines. \textit{Right panel:} Comparison between the inversion kernels in the continuum ($K^{cont}_{ij}$ for three of the \methane~lines in dot-dashed lines, and $K_{ij}$ for other wavenumbers in the continuum in dashed lines) and the sum of the inversion kernels $K^{CH_4}_{ij}$ of the \methane~rotational lines.   \methane~rotational lines dominate the temperature retrievals in the lower stratosphere, generally from 6 to 25~mbar  (and up to 35~mbar,  depending on the datasets). The continuum emission probes temperatures at pressures higher than 50~mbar, mainly in the troposphere.}
		\label{fig_cf}
	    \end{figure}

		The sensitivity of the spectra to the temperature can be measured with the inversion kernels for the temperature (defined as $K_{ij}~=~\frac{\partial I_i}{\partial T_j}$, where $I_i$ is the radiance measured at wavenumber $w_i$, and $T_j$ the temperature at pressure level $p_j$) for several wavenumbers. The contribution of the methane lines to the temperature measurement can be isolated by defining their own inversion kernels $K^{CH_4}_{ij}$ as follows:
	\begin{equation}
		K^{CH_4}_{ij} = K_{ij} - K^{cont}_{ij}
	\end{equation}
	\noindent where $K^{cont}_{ij}$ is the inversion kernel of the continuum for the same wavenumber. Figure \ref{fig_cf} shows $K^{CH_4}_{ij}$ for three of the rotational methane lines in the left panel, and the comparison between the sum of the 10 $K^{CH_4}_{ij}$ (for the 10 rotational \methane~lines) and inversion kernels for the continuum ($K^{cont}_{ij}$ at the wavenumbers of the \methane~lines and $K_{ij}$ outside of the \methane~lines) in the right panel. The \methane~lines allow us to measure lower stratospheric temperatures generally between 6~mbar and 25~mbar, with a maximal sensitivity at 15~mbar. The continuum emission mainly probes temperatures at higher pressures, around the tropopause and in the troposphere.  The continuum emission mostly originates from the $\mathrm{N_2}$-$\mathrm{N_2}$ and $\mathrm{N_2}$-\methane~collisions induced absorption with some contribution from the hazes, for which we have limited constraints.  However, Fig. \ref{fig_cf} shows that the continuum emission comes from pressure levels located several scale heights below the region probed by the \methane~lines, so the lack of constraints on the hazes and tropospheric temperatures does not affect the lower stratospheric temperatures which are the main focus of this study.\\ 
	
	\subsection{Error sources} 
		The main error sources in our temperature retrievals are the measurement noise and the uncertainties related to the retrieval process such as forward modelling errors or the smoothing of the temperature profile. The total error on the temperature retrieval is estimated by NEMESIS and is in the order of 2~K from 6~mbar to 25~mbar.\\ 
		
\begin{figure}
            \centering
            \includegraphics[width=1\columnwidth]{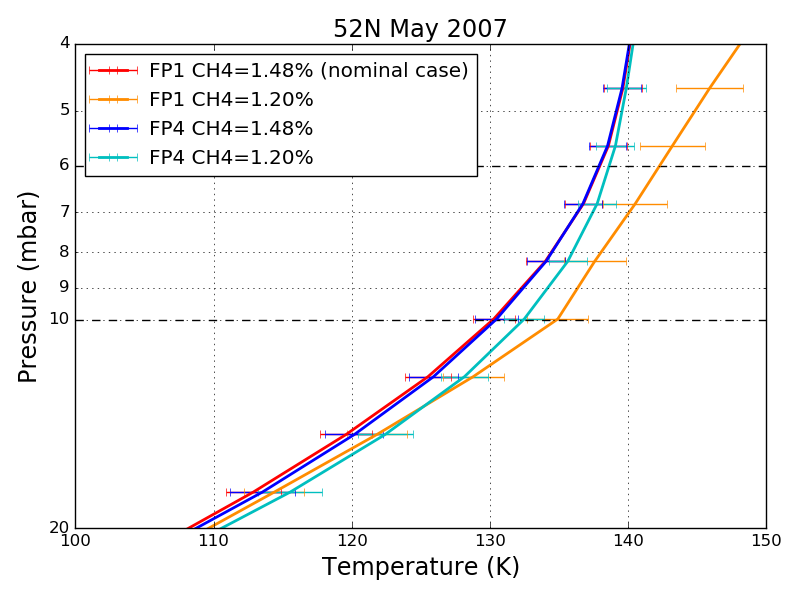}\\
            \includegraphics[width=1\columnwidth]{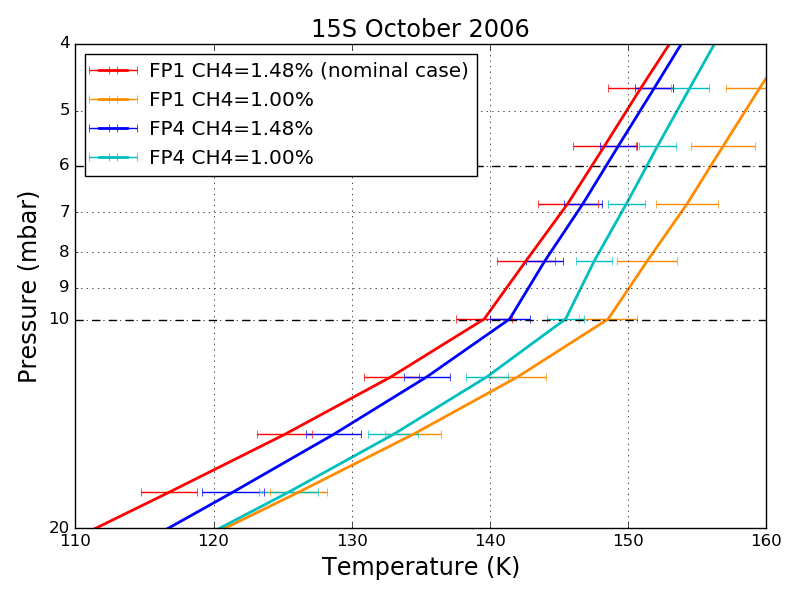}\\
            \caption{Temperature profiles from CIRS FP1 and FP4 nadir observations at $52^{\circ}$N in May 2007 (top panel) and $15^{\circ}$S in October 2006 (bottom panel), retrieved with the methane abundances measured by \citet{Niemann2010} (nominal value in this study) and \citet{Lellouch2014}. In both cases, the nominal value  from \citet{Niemann2010} yields a better agreement between the two types of observations.} 
            \label{fig_TCH4}
        \end{figure}

       The other possible error source is the uncertainty on \methane~abundance, as \citet{Lellouch2014} showed that it can vary from 1\% to 1.5\% at 15 mbar. We performed additional temperature retrievals on several datasets, in order to assess the effects of these variations on the temperature retrievals. First, we selected datasets for which \methane~abundance was measured by \citet{Lellouch2014}. In Figure \ref{fig_TCH4}, we show examples of these tests for two of these datasets: $52^{\circ}$N in May 2007 and $15^{\circ}$S in October 2006, for which \citet{Lellouch2014} measured respective \methane~abundances of $q_{CH_4} = 1.20 \pm 0.15\%$ and $q_{CH_4} = 0.95 \pm 0.08 \%$ (the nominal value for our retrievals is $q_{CH_4} = 1.48 \pm 0.09\%$ from \citet{Niemann2010}). At $52^{\circ}$N, the temperature profile obtained with the methane abundance from \citet{Lellouch2014} does not differ by more than 4~K from the nominal temperature profile. At 15 mbar (where the sensitivity to temperature is maximal in our retrievals), the difference of temperature between these two profiles is 2~K. Even a \methane~volume mixing ratio as low as 1\% yields a temperature only 4~K warmer than the nominal temperature at 15~mbar. At $15^{\circ}$S, the difference of temperature between the nominal retrieval and the retrieval with the methane abundance retrieved by \citet{Lellouch2014} ($q_{CH_4}=0.95\%$), is approximately 9~K on the whole pressure range.\\
       
       We performed additional temperature retrievals using CIRS FP4 nadir spectra measured at the same times and latitudes as the two datasets shown in Figure \ref{fig_TCH4}. In FP4 nadir spectra, the methane band $\nu_4$ is visible between $1200~\mathrm{cm^{-1}}$ and $1360~\mathrm{cm^{-1}}$. This spectral feature allows us to probe temperature between 0.1~mbar and 10~mbar, whereas methane rotational lines in the CIRS FP1 nadir spectra generally probe temperature between 6~mbar and 25~mbar. Temperature can thus be measured with both types of retrievals from 6~mbar to 10~mbar.  We performed FP4 temperature retrievals with the nominal methane abundance and the abundances measured by \citet{Lellouch2014}, as shown in Figure \ref{fig_TCH4}. FP4 temperature retrievals seem less sensitive to changes in the methane volume mixing ratio, as they yield a maximal temperature difference of 3~K at $52^{\circ}$N , and 4~K at $15^{\circ}$S between 6~mbar and 10~mbar. In both cases, FP1 and FP4 temperature retrievals are in better agreement in their common pressure range when the nominal methane abundance ($q_{CH_4}=1.48\%$) is used for both retrievals. This suggests that $q_{CH_4}=1.48\%$ is the best choice, at least in the pressure range covered by both types of temperature retrievals (from 6~mbar to 10 mbar). Changing the abundance of \methane~in the whole stratosphere seems to induce an error on the temperature measurements between 6~mbar and 10 mbar (up to 9~K at $15^{\circ}$S), which probably affects the temperature at 15~mbar in the FP1 retrievals, because of the vertical resolution of nadir retrievals (represented by the width of the inversion kernels in Fig. \ref{fig_cf}).  Consequently, assessing the effects of \methane~abundance variations on temperature at 15~mbar by changing $q_{CH_4}$ in the whole stratosphere seems to be a very unfavourable test, and the uncertainties on temperature determined by this method are probably overestimated for the FP1 temperature retrievals. Overall, when retrieving temperature from CIRS FP1 nadir spectra with $q_{CH_4}=1\%$ for datasets spanning different times and latitudes, we found temperatures warmer than our nominal temperatures by 2~K to 10~K at 15~mbar, with an average of 5~K. In \cite{Lellouch2014}, authors found that temperature changes by 4-5 K on the whole pressure range when varying $q_{CH_4}$ at $15^{\circ}$S, but they determined temperatures using FP4 nadir and limb data, which do not probe the 15 mbar pressure level.\\

\section{Results}
\label{sect_res}
	\begin{figure}[!hp]
		\includegraphics[width=1\columnwidth]{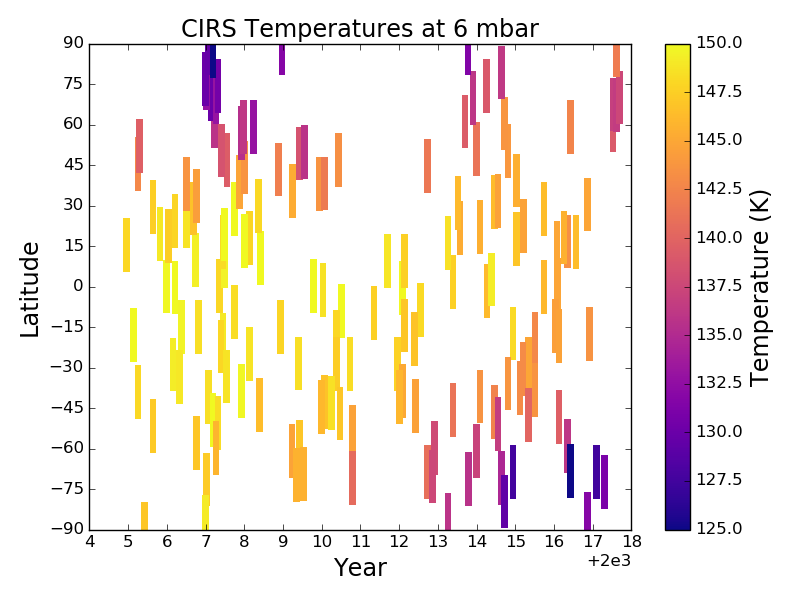}\\
		\includegraphics[width=1\columnwidth]{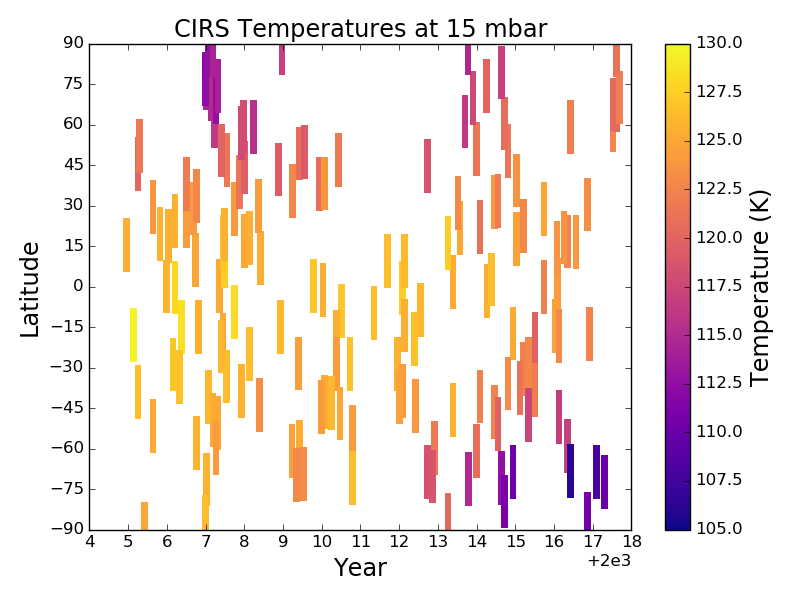}
		\caption{Evolution of temperatures at 6~mbar (120~km) and 15~mbar (85~km) from northern winter (2004) to summer (2017). The length of the markers shows the average size of the field of view of the CIRS FP1 detector. Temperatures exhibit similar strong seasonal changes at both pressure levels, especially at the poles.}
		\label{fig_ev_saiso}
	\end{figure}	
	
	\begin{figure}[!h]
		\includegraphics[width=1\columnwidth]{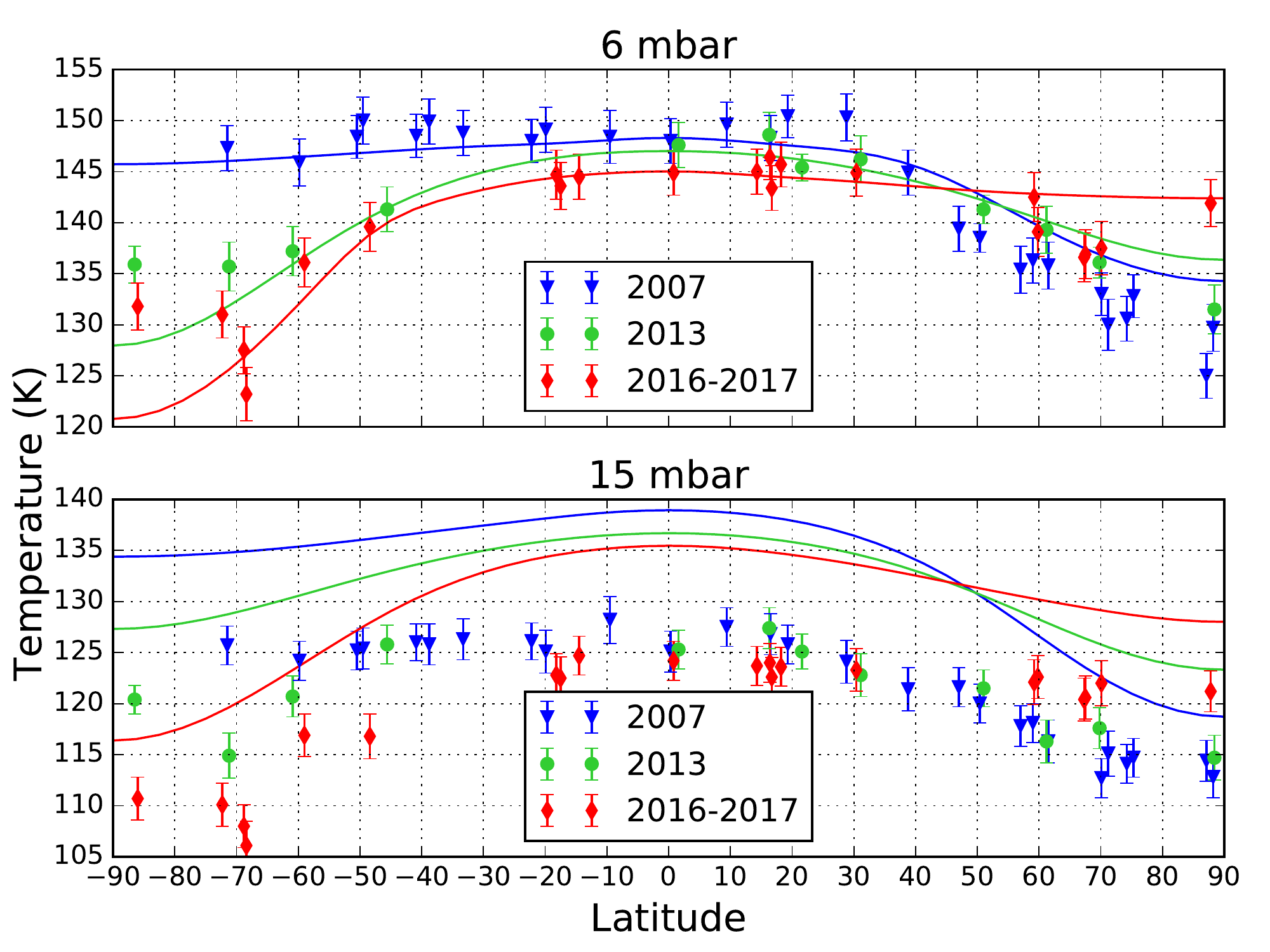}

		\caption{Meridional distribution of temperatures at 6~mbar (120~km) and 15~mbar (85~km), for three different seasons: late northern winter (2007, blue triangles), mid-spring (2013, green circles), and near summer solstice (from July 2016 to September 2017, red diamonds). The plain lines are the meridional distributions given by GCM simulations at comparable seasons (see section \ref{sect_discu}). In both observations and model the meridional gradient of temperatures evolves from one season to another at both pressure levels.}
		\label{fig_var_saiso}
	\end{figure}
			Figures \ref{fig_ev_saiso} and \ref{fig_var_saiso} show the temperatures measured with Cassini/CIRS far-IR nadir data at 6~mbar (minimal pressure probed by the CIRS far-IR nadir observations) and 15~mbar (pressure level where these observations are the most sensitive). Figure \ref{fig_ev_saiso} maps the seasonal evolution of temperatures throughout the Cassini mission (from 2004 to 2017, i.e. from mid-northern winter to early summer), while Figure \ref{fig_var_saiso} is focused on the evolution of the meridional gradient of temperature from one season to another. In both figures, both pressure levels exhibit significant seasonal variations of temperature and follow similar trends. Maximal temperatures are reached near the equator in 2005 (152~K at 6~mbar, 130~K at 15~mbar, at $18^{\circ}$S, at $L_S=300^{\circ}$), while the minimal temperatures are reached at high southern latitudes in autumn (123~K at 6~mbar, 106~K at 15~mbar at $70^{\circ}$S in 2016, at $L_S=79^{\circ}$).\\  
		
			The maximal seasonal variations of temperature are located at the poles for both pressure levels. At high northern latitudes ($60^\circ$N - $90^\circ$N), at 15~mbar, the temperature increased overall from winter to summer solstice. For instance at $70^{\circ}$N, temperature increased by 10~K from January 2007 to September 2017.  At 6~mbar, temperatures at  $60^{\circ}$N stayed approximately constant from winter to spring, whereas latitudes poleward from $70^{\circ}$N warmed up.  At $85^{\circ}$N, the temperature increased continuously from 125~K in March 2007 to 142~K in September 2017.\\
            
			In the meantime, at high southern latitudes ($60^\circ$S - $90^\circ$S), at 6~mbar and 15~mbar, temperatures strongly decreased from southern summer (2007) to late autumn (2016). It is the largest seasonal temperature change we measured in the lower stratosphere. At $70^{\circ}$S, temperature decreased by 24~K at 6~mbar and by 19~K at 15~mbar between January 2007 and June 2016. This decrease seems to be followed by a temperature increase toward winter solstice. At $70^{\circ}$S, temperatures varied by $+8$~K at 6~mbar from June 2016 to April 2017. Temperatures at high southern latitudes began to evolve in November 2010 at 6~mbar, and 2 years later (in August 2012) at 15~mbar.\\
			
			Other latitudes experience moderate seasonal temperature variations. At low latitudes (between $30^{\circ}$N and $30^{\circ}$S), temperature decreased overall from 2004 to 2017 at both pressure levels. For instance, at the equator, at 6~mbar temperature decreased by 6~K from 2006 to 2016.  At mid-southern latitudes, temperatures stayed constant from summer (2005) to mid-autumn (June 2012 at 6~mbar,  and May 2013 at 15~mbar), then they decreased by approximately 10~K from 2012-2013 to 2016. At mid-northern latitudes temperatures increased overall from winter to spring. At $50^{\circ}$N, temperature increased from 139~K to 144~K from 2005 to 2014.  In Figure \ref{fig_var_saiso}, at 6~mbar and 15~mbar, the meridional temperature gradient evolves from one season to another. During late northern winter, temperatures were approximately constant from $70^{\circ}$S to $30^{\circ}$N, and then decreased toward the North pole. In mid-spring, temperatures were  decreasing from equator to poles. Near the summer solstice, at 15~mbar, the meridional temperature gradient reversed compared to winter (summer temperatures constant in northern and low southern latitudes then decreasing toward the South Pole), while at 6~mbar,  temperatures globally decrease from the equator to the South pole and $70^{\circ}$N, then increase slightly between $70^{\circ}$N and $90^{\circ}$N. At 15~mbar, most of these changes in the shape of the temperature distribution occur because of the  temperature variations poleward from $60^{\circ}$. At 6~mbar, temperature variations occur mostly in the southern hemisphere at latitudes higher than $40^{\circ}$S, and near the North pole at latitudes higher than $70^{\circ}$N.\\  
	
	\begin{figure}[!h]
		\includegraphics[width=1\columnwidth]{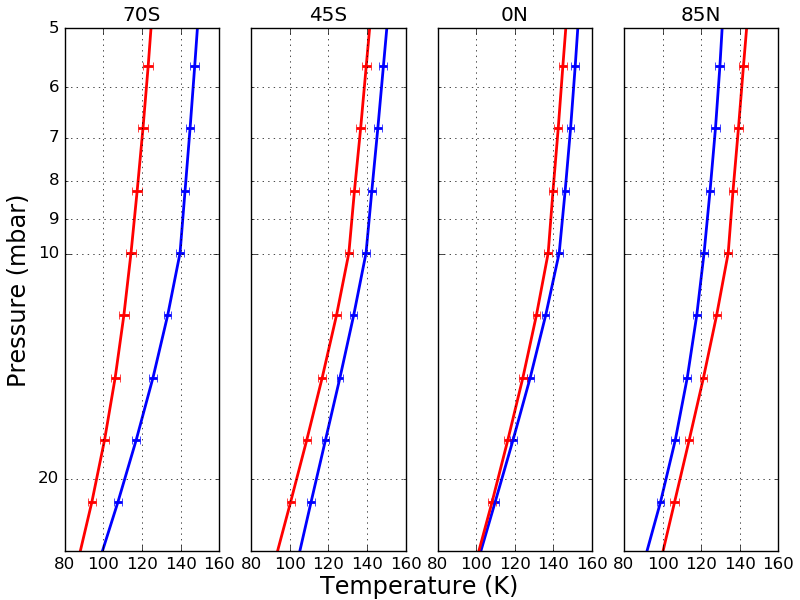}
		\caption{Temperature variations in the lower stratosphere during the Cassini mission for different latitudes. The blue profiles were measured during northern winter (in 2007). The red profiles were measured in late northern spring (in 2017 for $85^{\circ}$N, in 2016 for the other latitudes). The seasonal temperature variations are observed at most latitudes, and on the whole probed pressure range.}
		\label{fig_grad_saiso_vert}
		\end{figure}
	
		Figure \ref{fig_grad_saiso_vert} shows the first and the last temperature profiles measured with CIRS nadir far-IR data, for several latitudes. As in Fig. \ref{fig_ev_saiso}, the maximal temperature variations are measured at high southern latitudes for all pressure levels. At $70^{\circ}$S, the temperature decreased by 25~K at 10~mbar. Below 10~mbar the seasonal temperature difference decreases rapidly with increasing pressure until it reaches 10~K at 25~mbar, whereas it is  nearly constant between 5~mbar and 10~mbar. $85^{\circ}$N also exhibits a decrease of the seasonal temperature gradient below the 10~mbar pressure level, although it is less pronounced than near the South pole. At $45^{\circ}$S, the temperature decreased by approximately 10~K from 2007 to 2016, over the whole probed pressure range. At the equator, the temperature varies by -5~K from 2005 to 2016 at 6~mbar and the amplitude of this variation seems to decrease slightly with increasing pressure until it becomes negligible at 25~mbar. However the amplitude of these variations is in the same range as the uncertainty on temperature due to potential \methane~variations.    \\

\section{Discussion}
\label{sect_discu}
			\subsection{Comparison with previous results}
	
		\begin{figure}[!h]
			\includegraphics[width=1\columnwidth]{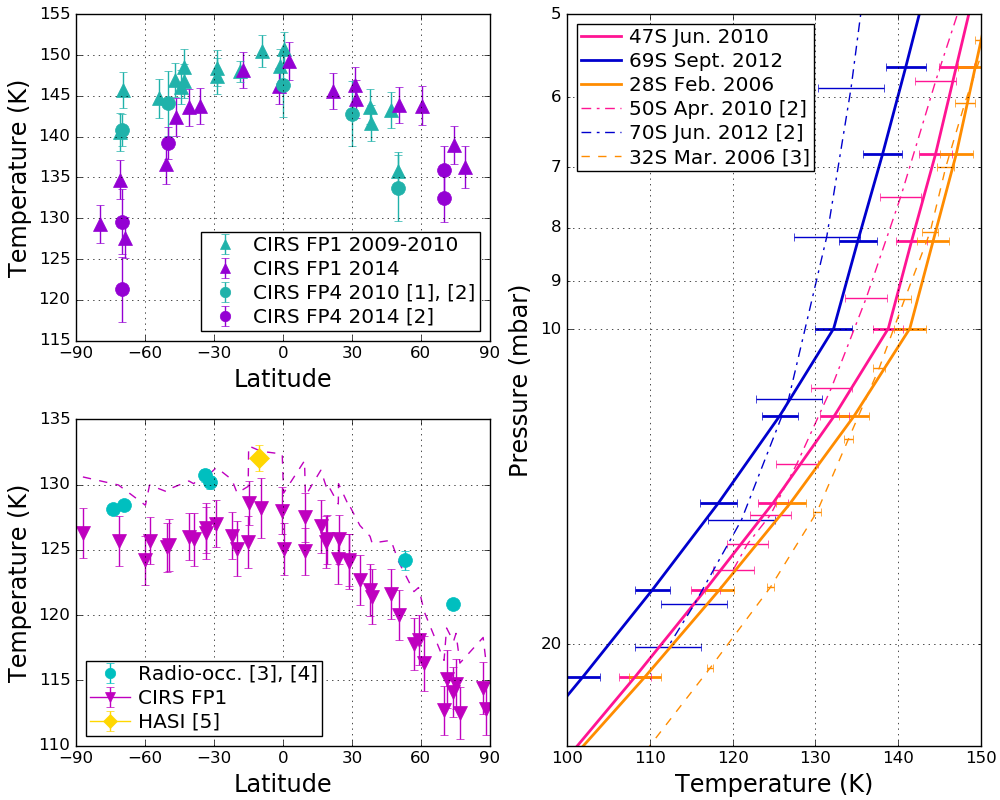}
			\caption{Comparison of nadir FP1 temperatures with previous studies.
				\textit{Top left panel:} Comparison between CIRS nadir FP1 (triangles) and CIRS nadir FP4 temperatures at 6~mbar (circles, \citet{Bampasidis2012}[1], and \citet{Coustenis2016}[2]) in 2010 (cyan) and 2014 (purple).
				\textit{Right panel:} Comparison between temperature profiles from CIRS nadir FP1 observations (thick solid lines), CIRS nadir FP4 observations (thin dot-dashed lines, \citet{Coustenis2016}[2]), and Cassini radio-occultation (thin dashed line, \citet{SchinderFlasarMaroufEtAl2011}[3]). Our results are in good agreement with CIRS FP4 temperatures, but diverge somewhat from radio-occultation profiles with increasing pressure. 
				\textit{Bottom left panel:} Comparison between temperatures at 15~mbar from our CIRS FP1 nadir measurements (magenta triangles), Cassini radio-occultations in 2006 and 2007 (cyan circles, \citet{SchinderFlasarMaroufEtAl2011,Schinder2012}, [3], [4]), and the Huygens/HASI measurement in 2005 (yellow diamond, \citet{Fulchignoni2005}, [5]).The dashed magenta line shows the potential effect of the \methane~variations observed by \citet{Lellouch2014}. If we take into account this effect, the agreement between our data, the radio-occultations and the HASI measurements is good.}
			\label{fig_prev_studies}
		\end{figure}	
			Figure \ref{fig_prev_studies} shows a comparison between our results and previous studies where temperatures have been measured in the lower stratosphere at similar epochs, latitudes and pressure levels. In the top left and right panels, our temperature measurements are compared to results from CIRS FP4 nadir observations \citep{Bampasidis2012, Coustenis2016} which probe mainly the 0.1-10~mbar pressure range. In the top left panel, the temperatures measured at 6~mbar by these two types of observations are in good agreement for the two considered epochs (2009-2010 and 2014). We obtain similar meridional gradients with both types of observations, even if FP4 temperatures are obtained from averages of spectra over bins of $10^{\circ}$ of latitudes (except at $70^{\circ}$N and $70^{\circ}$S where the bins are $20^{\circ}$ wide in latitude), whereas the average size in latitude of the field of view of the FP1 detector is $20^{\circ}$. It thus seems than the wider latitudinal size of the FP1 field of view has little effect on our temperature measurements. In the right panel, our temperature profiles are compared to two profiles measured by \citet{Coustenis2016} using CIRS FP4 nadir observations (at $50^{\circ}$S in April 2010, and at $70^{\circ}$S in June 2012), and with Cassini radio-occultations measurements from \citet{SchinderFlasarMaroufEtAl2011,Schinder2012}, which probe the atmosphere from the surface to 0.1~mbar (0 - 300~km). CIRS FP1 and FP4 temperature profiles are in good averall agreement. The profile we measured at $28^{\circ}$S in February 2006 and the corresponding radio-occultation profile are within error bars for pressures lower than 13~mbar, then the difference between them increases up to 8~K at 25~mbar. The bottom left panel of Fig. \ref{fig_prev_studies} shows the radio-occultation temperatures in 2006 and 2007 compared to CIRS nadir FP1 temperatures at 15~mbar, where their sensitivity to the temperature is maximal. Although, the radio-occultations temperatures are systematically higher than the CIRS temperatures by 2~K to 6~K, they follow the same meridional trend. CIRS FP1 temperatures at the equator are also lower than the temperature measured by the HASI instrument at 15~mbar during Huygens descent in Titan's atmosphere in 2005. If we take into account the effect of the spatial variations of \methane~at 15~mbar observed by \citet{Lellouch2014} by decreasing the \methane~abundance to 1\% (the lower limit in \citet{Lellouch2014}) in the CIRS FP1 temperature measurements (dashed line in the middle panel of Fig. \ref{fig_prev_studies}), the agreement between the three types of observations is good in the southern hemisphere. The differences between radio-occultations, HASI and CIRS temperatures might also be explained by the difference of vertical resolution. Indeed nadir observations have a vertical resolution in the order of 50~km while radio-occultations and HASI observations have respective vertical resolutions of 1~km and 200~m around 15~mbar.\\     

\subsection{Effects of Saturn's eccentricity}		
			\begin{figure}[!h]
			\includegraphics[width=1\columnwidth]{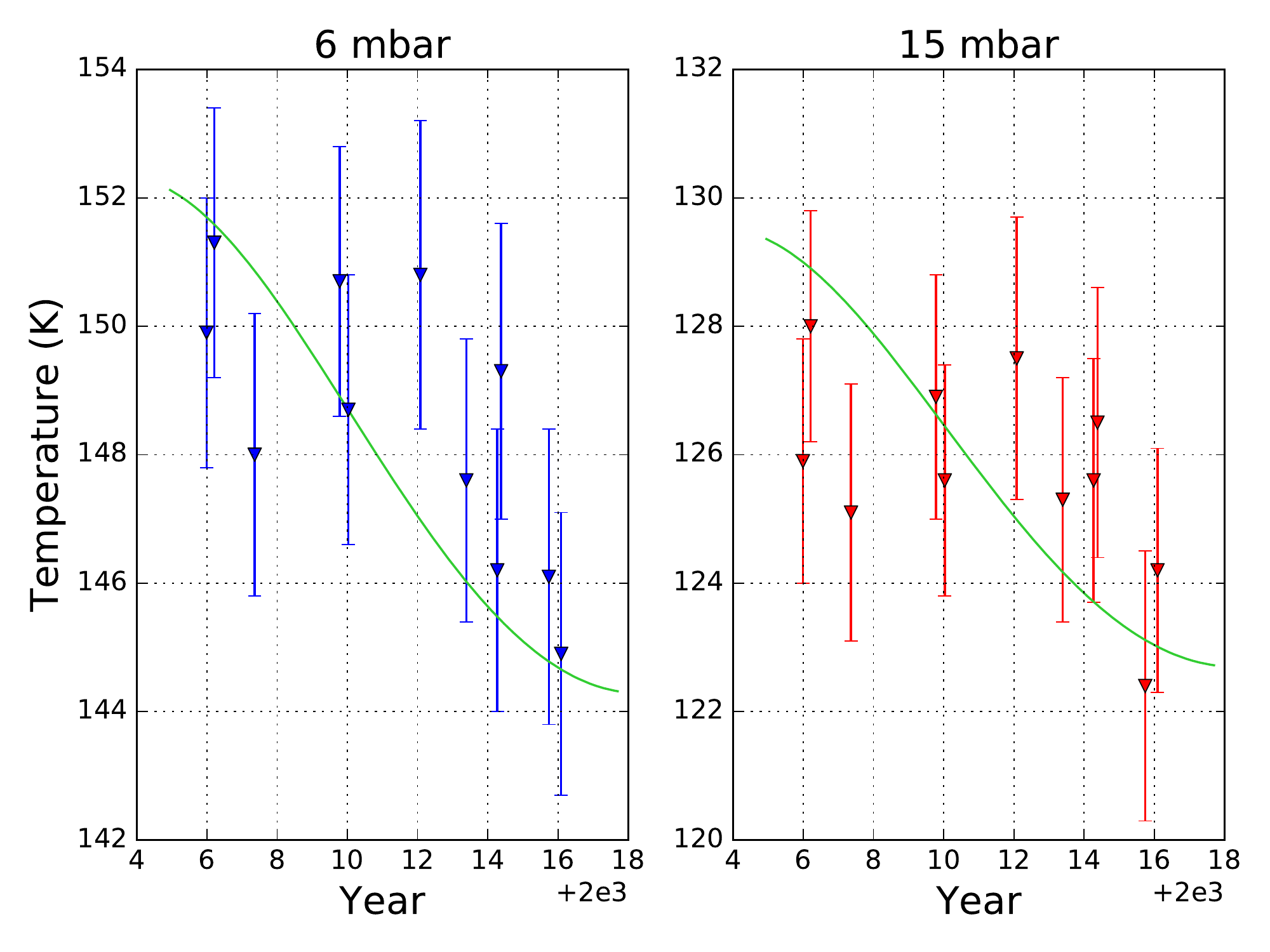}
			\caption{Temporal evolution of Titan's lower stratospheric temperatures at the equator ($5^{\circ}$N - $5^{\circ}$S) at 6~mbar (left panel) and 15~mbar (right panel), compared with a simple model of the evolution of the temperature as a function of the distance between Titan and the Sun (green line). The reduced $\chi^2$ between this model and the observations is 0.95 at 6~mbar and 1.07 at 15~mbar. The amplitude of the temperature variations at Titan's equator throughout the Cassini mission can be explained by the effect of Saturn's eccentricity.}  
			\label{fig_eccentricity}
		\end{figure}
		 Because of Saturn's orbital eccentricity of 0.0565, the distance between Titan and the Sun varies enough to affect significantly the insolation. For instance, throughout the Cassini mission, the solar flux received at the equator has decreased by 19\% because of the eccentricity.  We make a simple model of the evolution of the temperature $T$ at the equator as a function of the distance between Titan and the Sun. In this model we assume that the temperature $T$ at the considered pressure level and at a given time depends only on the absorbed solar flux $F$ and we neglect the radiative exchanges between atmospheric layers: 
		 
		 \begin{equation}
		 \epsilon \sigma T^4 = F
		 \end{equation} 
		 
		 \noindent where $\epsilon$ is the emissivity of the atmosphere at this pressure level, and $\sigma$ the Stefan-Boltzmann constant.  $T$ can thus be defined as a function of the distance $d$ between Titan and the Sun: 
		 
		 \begin{equation}
		 	T^4 = \frac{\alpha L_{\odot}}{16\epsilon\sigma\pi d^2}
		 \label{eq_T_dist}
		 \end{equation}
		 
		 \noindent where $L_\odot$ is the solar power, and $\alpha$ the absorptivity of the atmosphere. If we choose a reference temperature $T_0$ where Titan is at a distance $d_0$ from the Sun, a relation similar  to (\ref{eq_T_dist}) can be written for $T_0$. If we assume $\epsilon$ and $\alpha$ to be constant, $T$ can then be written as:
		 
		 \begin{equation}
		 	T = T_0 \sqrt{\frac{d_0}{d}}
		 \end{equation}
		   
		 Figure \ref{fig_eccentricity} shows a comparison between this model and the temperatures measured between $5^{\circ}$N and $5^{\circ}$S from 2006 to 2016, at 6~mbar and 15~mbar. We choose $T_0$ as the temperature at the beginning of the observations (December 2005/January 2006) which provides the best fit between our model and the observations while being consistent with the observations at the same epoch ($T_0=151.7$~K at 6~mbar, and $T_0=129$~K at 15~mbar). At 6~mbar, we measure a temperature decrease from 2006 to 2016. This is similar to what has been measured at 4~mbar by \citet{Bezard2018} with CIRS mid-IR observations, whereas their radiative-dynamical model predicts a small temperature maximum around the northern spring equinox (2009). At 15~mbar, equatorial temperatures are mostly constant from 2005 to 2016, with a marginal decrease in 2016. Our model predicts temperature variations of 8~K at 6~mbar and 7~K at 15~mbar from 2006 to 2016. Both predictions are consistent with the measurements and with radiative timescales shorter than one Titan year at 6~mbar and 15~mbar, as in \citet{Bezard2018} where they are respectively equal to 0.024~Titan year and 0.06~Titan year. At both pressure levels,  the model captures the magnitude of the temperature change, but does not fully match its timing or shape (especially in 2012-2014), implying that a more sophisticated model is needed. The remaining differences between our model and the temperature measurements could be decreased by adding a temporal lag to our model (2-3~years at 6~mbar and 3-4~years at 15~mbar), but the error bars on the temperature measurements are too large to constrain the lag to a value statistically distinct from zero.  Even with this potential lag, the agreement between the model and the temperatures measured at 6~mbar shows that the amplitude of the temporal evolution throughout the Cassini mission may be explained by the effects of Saturn's eccentricity. At 15~mbar, given the error bars and the lack of further far-IR temperature measurements at the equator in 2016 and 2017, it remains difficult to draw a definitive conclusion about the influence of Saturn's eccentricity at this pressure level.\\ 
	
	\subsection{Implication for radiative and dynamical processes of the lower stratosphere}
	
		In  Section \ref{sect_res}, we showed that in the lower stratosphere, the seasonal evolution of the temperature is maximal at high latitudes, especially at the South Pole. At 15~mbar, the strong cooling of high southern latitudes started in 2012, simultaneously with the increase in \cyano, \diacety, and \methyla~abundances measured at the same latitudes and pressure-level in \citet{Sylvestre2018}. We also show that this cooling affects the atmosphere at least down to the 25~mbar pressure level (altitude of 70~km). The enrichment of the gases and cooling are consistent with the onset of a subsidence above the South Pole during autumn, as predicted by GCMs \citep{Newman2011, Lebonnois2012a}, and inferred from previous CIRS observations at higher altitudes \citep{Teanby2012, Vinatier2015, Coustenis2016}. As Titan's atmospheric circulation transitions from two equator-to-poles cells (with upwelling above the equator and subsidence above the poles) to a single pole-to-pole cell (with a descending branch above South Pole), this subsidence drags downward photochemical species created at higher altitudes toward the lower stratosphere. \citet{Teanby2017} showed that enrichment in  trace gases may be so strong that their cooling effect combined with the insolation decrease may exceed the adiabatic heating between 0.3~mbar and 10~mbar (100 - 250~km). Our observations show that this phenomena may be at play down as deep as 25~mbar.\\

\begin{figure}[!h]
			\includegraphics[width=1\columnwidth]{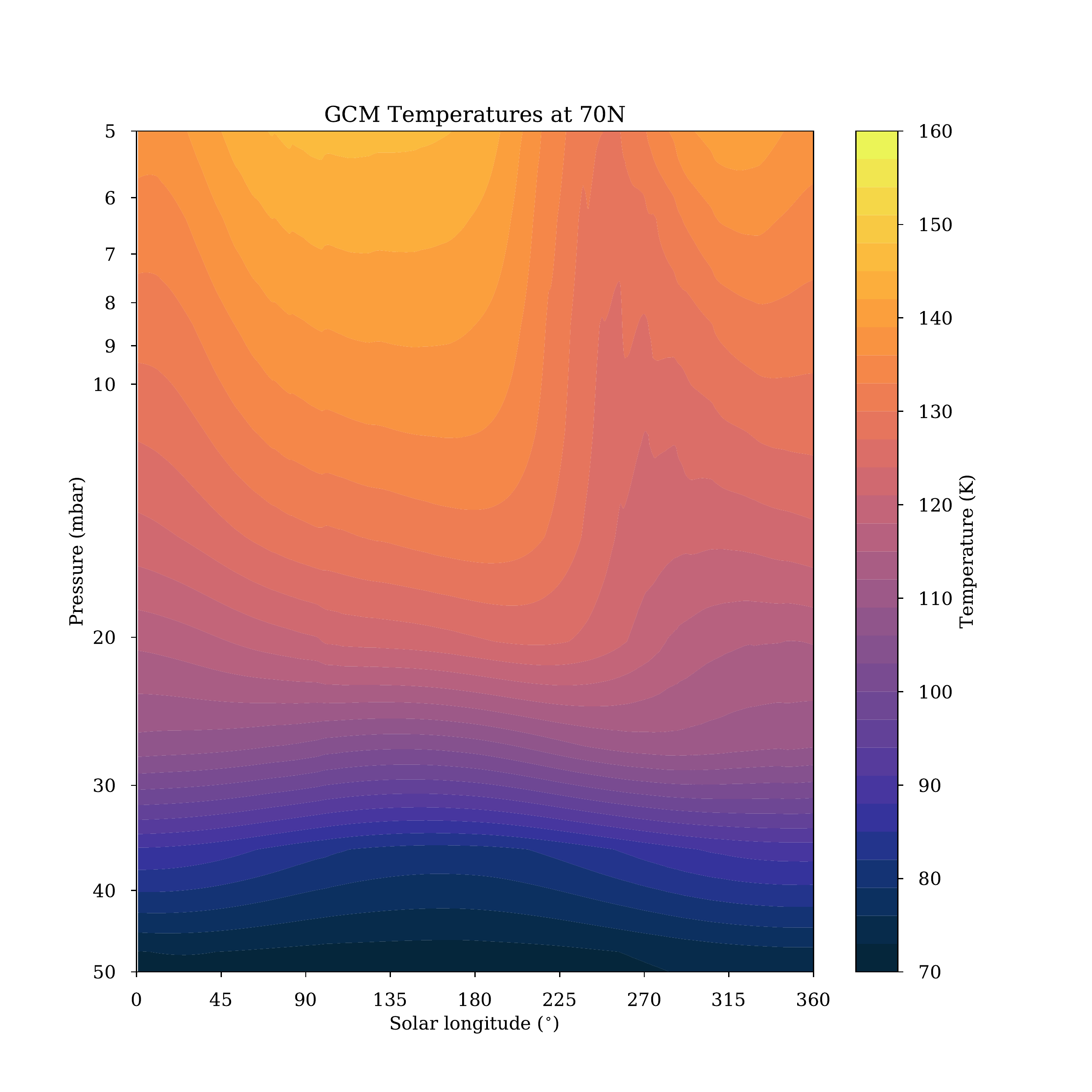}
			\caption{Seasonal evolution of Titan's lower stratospheric temperatures modeled by the IPSL 3D-GCM at 70$^{\circ}$N - between 5~mbar and 50~mbar, starting at northern spring equinox. In the pressure range probed by the CIRS far-IR observations (from 6~mbar to 25~mbar), there is a strong asymmetry between the rapid temperature changes after autumn equinox ($L_S = 180^{\circ}$) and the slow evolution of the thermal structure after spring equinox ($L_S = 0^{\circ}$). }
			\label{fig_map_temp_gcm70N}
		\end{figure}

        We compare retrieved temperature fields with results of simulations from IPSL 3D-GCM \citep{Lebonnois2012a} with an updated radiative transfer scheme \citep{Vatantd'Ollone2017} now based on a flexible \textit{correlated-k} method and up-to-date gas spectroscopic data \citep{Rothman2013}. It does not take into account the radiative feedback of the enrichment in hazes and trace gases in the polar regions, but it nevertheless appears that there is a good agreement in terms of seasonal cycle between the model and the observations. As shown in Figure \ref{fig_var_saiso}, at 6~mbar meridional distributions and values of temperatures in the model match well the observations. It can be pointed out that in both model and observations there is a noticeable asymmetry between high southern latitudes where the temperature decreases rapidly from the equinox to winter, and high northern latitudes which evolve more slowly from winter to summer. For instance, in both CIRS data and model, between 2007 and 2013 at 6~mbar and $70^{\circ}$N the atmosphere has warmed by only about 2~K, while in the meantime at $70^{\circ}$S it has cooled by about 10-15~K. This is consistent with an increase of radiative timescales at high northern latitudes (due to lower temperatures, \citet{Achterberg2011}) which would remain cold for approximately one season even after the return of sunlight. Figure \ref{fig_map_temp_gcm70N} shows the temporal evolution of the temperature at $70^{\circ}$N over one Titan year in the lower stratosphere in the GCM simulations and also emphasizes this asymmetry between the ingress and egress of winter at high latitudes. In Figure \ref{fig_var_saiso}, at 15~mbar modeled temperatures underestimate the observations by roughly 5-10~K, certainly due to a lack of infrared coolers such as clouds condensates \citep{Jennings2015}. However, observations and simulations exhibit similar meridional temperature gradients for the three studied epochs, and similar seasonal temperature evolution. For instance, in 2016-2017 we measured a temperature gradient of -11~K between the North and South Pole, whereas GCM simulations predict a temperature gradient of -12~K. At $70^{\circ}$S, temperature decreases by 10~K between 2007 and 2016-2017 in the GCM and in our observations. Besides, at 15~mbar, the seasonal behaviour remains the same as at 6~mbar, although more damped. Indeed comparison with GCM results also supports the idea that the seasonal effects due to the variations of insolation are damped with increasing depth in the lower stratosphere and ultimately muted below 25 mbar, as displayed in Figure \ref{fig_map_temp_gcm70N}. At lower altitudes the seasonal cycle of temperature at high latitudes is even inverted with temperatures increasing in the winter and decreasing in summer. Indeed at these altitudes, due to the radiative timescales exceeding one Titan year, temperature is no more sensitive to the seasonal variations of solar forcing, but to the interplay of ascending and descending large scale vertical motions of the pole-to-pole cell, inducing respectively adiabatic heating above winter pole and cooling above summer pole, as previously discussed in \citet{Lebonnois2012a}.

Further analysis of simulations – not presented here - also show that after 2016, temperatures at high southern latitudes began to slightly increase again at 6~mbar, which is consistent with observations, whereas at 15~mbar no change in the trend is observed, certainly due to a phase shift of the seasonal cycle between the two altitudes induced by the difference of radiative timescales, which is also illustrated in Figure \ref{fig_map_temp_gcm70N}.  \\

		We  also show in Figure \ref{fig_grad_saiso_vert} that at high southern latitudes, from 6 to 10~mbar seasonal temperature variations are approximately constant with pressure and can be larger than 10~K, whereas they decrease with increasing pressure below 10~mbar. This transition at 10~mbar may be caused by the increase of radiative timescales in the lower stratosphere. \citet{Strobel2010} estimated that the radiative timescale increases from one Titan season at 6~mbar to half a Titan year at 12~mbar. It can thus be expected that this region should be a transition zone between regions of the atmosphere where the atmospheric response to the seasonal insolation variations is significant and comes with little lag, to regions of the atmosphere where they are negligible.  However, this transition should be observable at other latitudes such as $45^{\circ}$S, whereas Figure \ref{fig_grad_saiso_vert} shows a seasonal gradient constant with pressure at this latitude. Furthermore, in \citet{Bezard2018}, the authors show that the method used to estimate radiative timescales in \citet{Strobel2010} tends to overestimate them, and that in their model radiative timescales are less than a Titan season down to the 35~mbar pressure level, which is more consistent with the seasonal variations measured at $45^{\circ}$S.\\
       
        The 10~mbar transition can also be caused by the interplay between photochemical, radiative and dynamical processes at high latitudes. Indeed, as photochemical species are transported downward by the subsidence above the autumn/winter pole, build up and cool strongly the lower atmosphere, the condensation level of species such as HCN, $\mathrm{HC_3N}$, $\mathrm{C_4H_2}$ or $\mathrm{C_6H_6}$ may be shifted upward, toward the 10~mbar level. Hence, below this pressure level, the volume mixing ratios of these gases would rapidly decrease, along with their cooling effect. Many observations, especially during the Cassini mission showed that during winter and autumn, polar regions host clouds composed of ices of photochemical species. For instance, the "haystack" feature showed in Fig. \ref{fig_spec} has been studied at both poles in  \citet{Coustenis1999,Jennings2012,Jennings2015},  and is attributed to a mixture of condensates, possibly of nitrile origin. Moreover, HCN ice has been measured in the southern polar cloud observed by \citet{deKok2014} with Cassini/VIMS observations. $\mathrm{C_6H_6}$ ice has also been detected by \citet{Vinatier2018} in CIRS observations of the South Pole. The condensation curve for \diacety~in \citet{Barth2017} is also consistent with the formation of  \diacety~ice around 10~mbar with the temperatures we measured at $70^{\circ}$S in 2016. These organic ices may also have a cooling effect themselves as \citet{Bezard2018} showed that at 9~mbar, the nitrile haze measured by \citet{Anderson2011} contributes to the cooling with an intensity comparable to the contribution of  gases such as $\mathrm{C_2H_2}$ and $\mathrm{C_2H_6}$.  \\  
		
\section{Conclusion}

	In this paper, we analysed all the available nadir far-IR CIRS observations to measure Titan's lower stratospheric temperatures (6~mbar - 25~mbar) throughout the 13 years of the Cassini mission, from northern winter to summer solstice. In this pressure range, significant temperature changes occur from one season to another. Temperatures evolve moderately at low and mid-latitudes (less than 10~K between 6 and 15~mbar).  At the equator, at 6~mbar we measure a temperature decrease mostly due to Saturn's eccentricity. Seasonal temperature changes are maximal at high latitudes, especially in the southern hemisphere where they reach up to -19~K at $70^{\circ}$S between summer (2007) and late autumn (2016) at 15~mbar. The strong seasonal evolution of high southern latitudes is due to a complex interplay between photochemistry, atmospheric dynamics  with the downwelling above the autumn/winter poles, radiative processes with a large contribution of the gases transported toward the lower stratosphere, and possibly condensation due to the cold autumn polar temperatures and strong enrichments in trace gases.\\
	    
Recent GCM simulations show a good agreement with the observed seasonal variations in this pressure range, even though these simulations do not include coupling with variations of opacity sources. In particular at high latitudes, the fast decrease of temperatures when entering winter and slower increase when getting into summer is well reproduced in these simulations.

\section*{Acknowledgements}
	This research was funded by the UK Sciences and Technology Facilities Research council (grant number ST/MOO7715/1) and the Cassini project. JVO and SL acknowledge support from the Centre National d'Etudes Spatiales (CNES). GCM simulations have been performed thanks to computation facilities provided by the Grand Équipement National de Calcul Intensif (GENCI) on the \textit{Occigen/CINES} cluster (allocation A0040110391). 
This research made use of Astropy, a community-developed core Python package for Astronomy \citep{2013A&A...558A..33A}, and matplotlib, a Python library for publication quality graphics \citep{Hunter:2007}

\section*{Appendix. Cassini/CIRS Datasets analysed in this study}	%

	\onecolumn
	\begin{longtable}[!h]{lcccc}
	\caption{\label{table_obs}Far-IR CIRS datasets presented in this study. N stands for the number of spectra measured during the acquisition. FOV is the field of view. The asterisk denotes datasets where two different latitudes were observed. }\\
	\hline
	\hline
	 Observations & Date & N & Latitude ($^{\circ}$N) & FOV ($^{\circ}$)\\
	\hline
		
		CIRS\_00BTI\_FIRNADCMP001\_PRIME & 12 Dec. 2004 &  224 & 16.4 & 20.3\\
		CIRS\_003TI\_FIRNADCMP002\_PRIME & 15 Feb. 2005 &  180 & -18.7 & 18.5\\
		CIRS\_005TI\_FIRNADCMP002\_PRIME & 31 Mar. 2005 &  241 & -41.1 & 25.7\\
		CIRS\_005TI\_FIRNADCMP003\_PRIME & 01 Apr. 2005 &  240 & 47.8 & 28.5\\
		CIRS\_006TI\_FIRNADCMP002\_PRIME & 16 Apr. 2005 &  178 & 54.7 & 29.9\\	
		CIRS\_009TI\_COMPMAP002\_PRIME & 06 Jun. 2005 &  184 & -89.7 & 21.1\\		
		CIRS\_013TI\_FIRNADCMP003\_PRIME & 21 Aug. 2005 &  192 & 30.1 & 15.5\\
		CIRS\_013TI\_FIRNADCMP004\_PRIME & 22 Aug. 2005 &  248 & -53.7 & 25.0\\
		CIRS\_017TI\_FIRNADCMP003\_PRIME & 28 Oct. 2005 &  119 & 20.1 & 19.8\\
		CIRS\_019TI\_FIRNADCMP002\_PRIME & 26 Dec. 2005 &  124 & -0.0 & 17.6\\
		CIRS\_020TI\_FIRNADCMP002\_PRIME & 14 Jan. 2006 &  107 & 19.5 & 19.7\\
		CIRS\_021TI\_FIRNADCMP002\_PRIME & 27 Feb. 2006 &  213 & -30.2 & 22.5\\
		CIRS\_022TI\_FIRNADCMP003\_PRIME & 18 Mar. 2006 &  401 & -0.4 & 18.4\\
		CIRS\_022TI\_FIRNADCMP008\_PRIME & 19 Mar. 2006 &  83 & 25.3 & 24.1\\
		CIRS\_023TI\_FIRNADCMP002\_PRIME & 01 May 2006 &  215 & -35.0 & 27.8\\
		CIRS\_024TI\_FIRNADCMP003\_PRIME & 19 May 2006 &  350 & -15.5 & 21.6\\
		CIRS\_025TI\_FIRNADCMP002\_PRIME & 02 Jul. 2006 &  307 & 25.1 & 21.7\\
		CIRS\_025TI\_FIRNADCMP003\_PRIME & 01 Jul. 2006 &  190 & 39.7 & 25.6\\
		CIRS\_028TI\_FIRNADCMP003\_PRIME & 07 Sep. 2006 &  350 & 29.7 & 19.7\\
		CIRS\_029TI\_FIRNADCMP003\_PRIME & 23 Sep. 2006 &  312 & 9.5 & 19.4\\
		CIRS\_030TI\_FIRNADCMP002\_PRIME & 10 Oct. 2006 &  340 & -59.1 & 23.4\\
		CIRS\_030TI\_FIRNADCMP003\_PRIME & 09 Oct. 2006 &  286 & 33.9 & 19.9\\
		CIRS\_031TI\_COMPMAP001\_VIMS & 25 Oct. 2006 &  160 & -14.5 & 16.3\\
		CIRS\_036TI\_FIRNADCMP002\_PRIME & 28 Dec. 2006 &  136 & -89.1 & 12.6\\
		CIRS\_036TI\_FIRNADCMP003\_PRIME & 27 Dec. 2006 &  321 & 78.6 & 21.0\\
		CIRS\_037TI\_FIRNADCMP001\_PRIME & 12 Jan. 2007 &  161 & 75.2 & 19.1\\
		CIRS\_037TI\_FIRNADCMP002\_PRIME & 13 Jan. 2007 &  107 & -70.3 & 20.6\\
		CIRS\_038TI\_FIRNADCMP001\_PRIME & 28 Jan. 2007 &  254 & 86.3 & 16.7\\
		CIRS\_038TI\_FIRNADCMP002\_PRIME & 29 Jan. 2007 &  254 & -39.7 & 22.0\\
		CIRS\_039TI\_FIRNADCMP002\_PRIME & 22 Feb. 2007 &  23 & 69.9 & 21.2\\
		CIRS\_040TI\_FIRNADCMP001\_PRIME & 09 Mar. 2007 &  159 & -49.2 & 21.1\\
		CIRS\_040TI\_FIRNADCMP002\_PRIME & 10 Mar. 2007 &  109 & 88.8 & 13.3\\
		CIRS\_041TI\_FIRNADCMP002\_PRIME & 26 Mar. 2007 &  102 & 61.2 & 19.3\\
		CIRS\_042TI\_FIRNADCMP001\_PRIME & 10 Apr. 2007 &  103 & -60.8 & 26.0\\
		CIRS\_042TI\_FIRNADCMP002\_PRIME & 11 Apr. 2007 &  272 & 71.5 & 22.6\\
		CIRS\_043TI\_FIRNADCMP001\_PRIME & 26 Apr. 2007 &  263 & -51.4 & 24.7\\
		CIRS\_043TI\_FIRNADCMP002\_PRIME & 27 Apr. 2007 &  104 & 77.1 & 20.0\\
		CIRS\_044TI\_FIRNADCMP002\_PRIME & 13 May 2007 &  104 & -0.5 & 18.8\\
		CIRS\_045TI\_FIRNADCMP001\_PRIME & 28 May 2007 &  231 & -22.3 & 22.6\\
		CIRS\_045TI\_FIRNADCMP002\_PRIME & 29 May 2007 &  346 & 52.4 & 29.5\\
		CIRS\_046TI\_FIRNADCMP001\_PRIME & 13 Jun. 2007 &  60 & 17.6 & 28.6\\
		CIRS\_046TI\_FIRNADCMP002\_PRIME & 14 Jun. 2007 &  102 & -20.8 & 19.0\\
		CIRS\_047TI\_FIRNADCMP001\_PRIME & 29 Jun. 2007 &  204 & 9.8 & 23.2\\
		CIRS\_047TI\_FIRNADCMP002\_PRIME & 30 Jun. 2007 &  238 & 20.1 & 23.7\\
		CIRS\_048TI\_FIRNADCMP001\_PRIME & 18 Jul. 2007 &  96 & -34.8 & 31.4\\
		CIRS\_048TI\_FIRNADCMP002\_PRIME & 19 Jul. 2007 &  260 & 49.5 & 35.8\\
		CIRS\_050TI\_FIRNADCMP001\_PRIME & 01 Oct. 2007 &  144 & -10.1 & 23.8\\
		CIRS\_050TI\_FIRNADCMP002\_PRIME & 02 Oct. 2007 &  106 & 29.9 & 19.7\\
		CIRS\_052TI\_FIRNADCMP002\_PRIME & 19 Nov. 2007 &  272 & 40.3 & 26.5\\
		CIRS\_053TI\_FIRNADCMP001\_PRIME & 04 Dec. 2007 &  223 & -40.2 & 25.8\\
		CIRS\_053TI\_FIRNADCMP002\_PRIME & 05 Dec. 2007 &  102 & 59.4 & 28.3\\
		CIRS\_054TI\_FIRNADCMP002\_PRIME & 21 Dec. 2007 &  107 & 60.4 & 21.1\\
		CIRS\_055TI\_FIRNADCMP001\_PRIME & 05 Jan. 2008 &  190 & 18.7 & 30.5\\
		CIRS\_055TI\_FIRNADCMP002\_PRIME & 06 Jan. 2008 &  284 & 44.6 & 22.2\\
		CIRS\_059TI\_FIRNADCMP001\_PRIME & 22 Feb. 2008 &  172 & -24.9 & 20.7\\
		CIRS\_059TI\_FIRNADCMP002\_PRIME & 23 Feb. 2008 &  98 & 17.1 & 20.0\\
		CIRS\_062TI\_FIRNADCMP002\_PRIME & 25 Mar. 2008 &  115 & 59.3 & 17.1\\
		CIRS\_067TI\_FIRNADCMP002\_PRIME & 12 May 2008 &  286 & 29.5 & 21.0\\
		CIRS\_069TI\_FIRNADCMP001\_PRIME & 27 May 2008 &  112 & -44.6 & 27.3\\
		CIRS\_069TI\_FIRNADCMP002\_PRIME & 28 May 2008 &  112 & 9.5 & 19.3\\
		CIRS\_093TI\_FIRNADCMP002\_PRIME & 20 Nov. 2008 &  161 & 43.7 & 21.1\\
		CIRS\_095TI\_FIRNADCMP001\_PRIME & 05 Dec. 2008 &  213 & -14.0 & 20.7\\
		CIRS\_097TI\_FIRNADCMP001\_PRIME & 20 Dec. 2008 &  231 & -10.9 & 23.7\\
		CIRS\_106TI\_FIRNADCMP001\_PRIME & 26 Mar. 2009 &  165 & -60.3 & 19.2\\
		CIRS\_107TI\_FIRNADCMP002\_PRIME & 27 Mar. 2009 &  164 & 33.5 & 30.4\\
		CIRS\_110TI\_FIRNADCMP001\_PRIME & 06 May 2009 &  282 & -68.1 & 25.7\\
		CIRS\_111TI\_FIRNADCMP002\_PRIME & 22 May 2009 &  168 & -27.1 & 23.1\\
		CIRS\_112TI\_FIRNADCMP001\_PRIME & 06 Jun. 2009 &  218 & 48.7 & 21.0\\
		CIRS\_112TI\_FIRNADCMP002\_PRIME & 07 Jun. 2009 &  274 & -58.9 & 20.2\\
		CIRS\_114TI\_FIRNADCMP001\_PRIME & 09 Jul. 2009 &  164 & -71.4 & 25.4\\
		CIRS\_115TI\_FIRNADCMP001\_PRIME & 24 Jul. 2009 &  146 & 50.7 & 20.1\\
		CIRS\_119TI\_FIRNADCMP002\_PRIME & 12 Oct. 2009 &  166 & 0.4 & 18.3\\
		CIRS\_122TI\_FIRNADCMP001\_PRIME & 11 Dec. 2009 &  212 & 39.8 & 24.7\\
		CIRS\_123TI\_FIRNADCMP002\_PRIME & 28 Dec. 2009 &  186 & -46.1 & 22.3\\
		CIRS\_124TI\_FIRNADCMP002\_PRIME & 13 Jan. 2010 &  272 & -1.2 & 19.0\\
		CIRS\_125TI\_FIRNADCMP001\_PRIME & 28 Jan. 2010 &  156 & 39.9 & 27.5\\
		CIRS\_125TI\_FIRNADCMP002\_PRIME & 29 Jan. 2010 &  280 & -44.9 & 27.3\\
		CIRS\_129TI\_FIRNADCMP001\_PRIME & 05 Apr. 2010 &  119 & -45.1 & 28.2\\
		CIRS\_131TI\_FIRNADCMP001\_PRIME & 19 May 2010 &  188 & -30.0 & 22.1\\
		CIRS\_131TI\_FIRNADCMP002\_PRIME & 20 May 2010 &  229 & -19.8 & 21.5\\
		CIRS\_132TI\_FIRNADCMP002\_PRIME & 05 Jun. 2010 &  167 & 49.4 & 27.4\\
		CIRS\_133TI\_FIRNADCMP001\_PRIME & 20 Jun. 2010 &  187 & -49.7 & 36.1\\
		CIRS\_134TI\_FIRNADCMP001\_PRIME & 06 Jul. 2010 &  251 & -10.0 & 20.0\\
		CIRS\_138TI\_FIRNADCMP001\_PRIME & 24 Sep. 2010 &  190 & -30.1 & 21.2\\
		CIRS\_139TI\_COMPMAP001\_PRIME* & 14 Oct. 2010 &  132 & -70.9 & 20.6\\
		CIRS\_139TI\_COMPMAP001\_PRIME* & 14 Oct. 2010 &  108 & -53.8 & 16.7\\
		CIRS\_148TI\_FIRNADCMP001\_PRIME & 08 May 2011 &  200 & -10.0 & 18.3\\
		CIRS\_153TI\_FIRNADCMP001\_PRIME & 11 Sep. 2011 &  227 & 9.9 & 19.0\\
		CIRS\_158TI\_FIRNADCMP501\_PRIME & 13 Dec. 2011 &  369 & -29.9 & 24.7\\
		CIRS\_159TI\_FIRNADCMP001\_PRIME & 02 Jan. 2012 &  275 & -42.2 & 23.7\\
		CIRS\_160TI\_FIRNADCMP001\_PRIME & 29 Jan. 2012 &  322 & -40.0 & 21.7\\
		CIRS\_160TI\_FIRNADCMP002\_PRIME & 30 Jan. 2012 &  280 & -0.2 & 18.3\\
		CIRS\_161TI\_FIRNADCMP001\_PRIME & 18 Feb. 2012 &  121 & 9.9 & 18.4\\
		CIRS\_161TI\_FIRNADCMP002\_PRIME & 19 Feb. 2012 &  89 & -15.0 & 17.3\\
		CIRS\_166TI\_FIRNADCMP001\_PRIME & 22 May 2012 &  318 & -19.9 & 19.9\\
		CIRS\_167TI\_FIRNADCMP002\_PRIME & 07 Jun. 2012 &  293 & -45.4 & 21.7\\
		CIRS\_169TI\_FIRNADCMP001\_PRIME & 24 Jul. 2012 &  258 & -9.7 & 20.7\\
		CIRS\_172TI\_FIRNADCMP001\_PRIME & 26 Sep. 2012 &  282 & 44.9 & 18.5\\
		CIRS\_172TI\_FIRNADCMP002\_PRIME & 26 Sep. 2012 &  270 & -70.4 & 23.2\\
		CIRS\_174TI\_FIRNADCMP002\_PRIME & 13 Nov. 2012 &  298 & -71.8 & 21.8\\
		CIRS\_175TI\_FIRNADCMP002\_PRIME & 29 Nov. 2012 &  299 & -59.9 & 19.3\\
		CIRS\_185TI\_FIRNADCMP001\_PRIME & 05 Apr. 2013 &  244 & 15.0 & 20.1\\
		CIRS\_185TI\_FIRNADCMP002\_PRIME & 06 Apr. 2013 &  303 & -88.9 & 16.8\\
		CIRS\_190TI\_FIRNADCMP001\_PRIME & 23 May 2013 &  224 & -0.2 & 25.6\\
		CIRS\_190TI\_FIRNADCMP002\_PRIME & 24 May 2013 &  298 & -45.0 & 20.0\\
		CIRS\_194TI\_FIRNADCMP001\_PRIME & 10 Jul. 2013 &  186 & 30.0 & 19.7\\
		CIRS\_195TI\_FIRNADCMP001\_PRIME & 25 Jul. 2013 &  186 & 19.6 & 24.5\\
		CIRS\_197TI\_FIRNADCMP001\_PRIME & 11 Sep. 2013 &  330 & 60.5 & 19.4\\
		CIRS\_198TI\_FIRNADCMP001\_PRIME & 13 Oct. 2013 &  187 & 88.9 & 8.7\\
		CIRS\_198TI\_FIRNADCMP002\_PRIME & 14 Oct. 2013 &  306 & -69.8 & 24.0\\
		CIRS\_199TI\_FIRNADCMP001\_PRIME & 30 Nov. 2013 &  329 & 68.4 & 23.9\\
		CIRS\_200TI\_FIRNADCMP001\_PRIME & 01 Jan. 2014 &  187 & 49.9 & 19.6\\
		CIRS\_200TI\_FIRNADCMP002\_PRIME & 02 Jan. 2014 &  210 & -59.8 & 21.3\\
		CIRS\_201TI\_FIRNADCMP001\_PRIME & 02 Feb. 2014 &  329 & 19.9 & 26.8\\
		CIRS\_201TI\_FIRNADCMP002\_PRIME & 03 Feb. 2014 &  234 & -39.6 & 20.9\\
		CIRS\_203TI\_FIRNADCMP001\_PRIME & 07 Apr. 2014 &  187 & 75.0 & 18.0\\
		CIRS\_203TI\_FIRNADCMP002\_PRIME & 07 Apr. 2014 &  239 & 0.5 & 27.5\\
		CIRS\_204TI\_FIRNADCMP002\_PRIME & 18 May 2014 &  199 & 0.4 & 27.0\\
		CIRS\_205TI\_FIRNADCMP001\_PRIME & 18 Jun. 2014 &  144 & -45.1 & 20.5\\
		CIRS\_205TI\_FIRNADCMP002\_PRIME & 18 Jun. 2014 &  161 & 30.3 & 19.1\\
		CIRS\_206TI\_FIRNADCMP001\_PRIME & 19 Jul. 2014 &  181 & -50.3 & 17.8\\
		CIRS\_206TI\_FIRNADCMP002\_PRIME & 20 Jul. 2014 &  161 & 30.6 & 18.4\\
		CIRS\_207TI\_FIRNADCMP001\_PRIME & 20 Aug. 2014 &  179 & -70.0 & 17.8\\
		CIRS\_207TI\_FIRNADCMP002\_PRIME & 21 Aug. 2014 &  163 & 79.7 & 17.6\\
		CIRS\_208TI\_FIRNADCMP001\_PRIME & 21 Sep. 2014 &  329 & -80.0 & 15.6\\
		CIRS\_208TI\_FIRNADCMP002\_PRIME & 22 Sep. 2014 &  175 & 60.5 & 17.8\\
		CIRS\_209TI\_FIRNADCMP001\_PRIME & 23 Oct. 2014 &  181 & -35.2 & 17.7\\
		CIRS\_209TI\_FIRNADCMP002\_PRIME & 24 Oct. 2014 &  233 & 50.5 & 18.5\\
		CIRS\_210TI\_FIRNADCMP001\_PRIME & 10 Dec. 2014 &  329 & -70.3 & 25.2\\
		CIRS\_210TI\_FIRNADCMP002\_PRIME & 11 Dec. 2014 &  237 & -19.6 & 27.6\\
		CIRS\_211TI\_FIRNADCMP001\_PRIME & 11 Jan. 2015 &  225 & 19.6 & 25.0\\
		CIRS\_211TI\_FIRNADCMP002\_PRIME & 12 Jan. 2015 &  258 & 40.0 & 19.3\\
		CIRS\_212TI\_FIRNADCMP002\_PRIME & 13 Feb. 2015 &  257 & -40.0 & 30.1\\
		CIRS\_213TI\_FIRNADCMP001\_PRIME & 16 Mar. 2015 &  187 & -31.6 & 19.6\\
		CIRS\_213TI\_FIRNADCMP002\_PRIME & 16 Mar. 2015 &  258 & 23.4 & 20.5\\
		CIRS\_215TI\_FIRNADCMP001\_PRIME & 07 May 2015 &  250 & -50.0 & 31.0\\
		CIRS\_215TI\_FIRNADCMP002\_PRIME & 08 May 2015 &  232 & -30.0 & 21.7\\
		CIRS\_218TI\_FIRNADCMP001\_PRIME & 06 Jul. 2015 &  249 & -20.0 & 19.9\\
		CIRS\_218TI\_FIRNADCMP002\_PRIME & 07 Jul. 2015 &  232 & -40.0 & 25.2\\
		CIRS\_222TI\_FIRNADCMP001\_PRIME & 28 Sep. 2015 &  125 & 30.0 & 21.7\\
		CIRS\_222TI\_FIRNADCMP002\_PRIME & 29 Sep. 2015 &  233 & -0.1 & 18.6\\
		CIRS\_230TI\_FIRNADCMP001\_PRIME & 15 Jan. 2016 &  282 & -15.0 & 19.5\\
		CIRS\_231TI\_FIRNADCMP001\_PRIME & 31 Jan. 2016 &  254 & 15.0 & 19.6\\
		CIRS\_231TI\_FIRNADCMP002\_PRIME & 01 Feb. 2016 &  236 & 0.4 & 18.9\\
		CIRS\_232TI\_FIRNADCMP001\_PRIME & 16 Feb. 2016 &  249 & -50.2 & 24.5\\
		CIRS\_232TI\_FIRNADCMP002\_PRIME & 17 Feb. 2016 &  92 & -19.8 & 21.5\\
		CIRS\_234TI\_FIRNADCMP001\_PRIME & 04 Apr. 2016 &  328 & 19.8 & 24.7\\
		CIRS\_235TI\_FIRNADCMP001\_PRIME & 06 May 2016 &  163 & -60.0 & 19.7\\
		CIRS\_235TI\_FIRNADCMP002\_PRIME & 07 May 2016 &  221 & 15.7 & 20.1\\
		CIRS\_236TI\_FIRNADCMP001\_PRIME & 07 Jun. 2016 &  88 & -70.5 & 20.5\\
		CIRS\_236TI\_FIRNADCMP002\_PRIME & 07 Jun. 2016 &  238 & 60.8 & 20.0\\
		CIRS\_238TI\_FIRNADCMP002\_PRIME & 25 Jul. 2016 &  220 & 15.4 & 20.5\\
		CIRS\_248TI\_FIRNADCMP001\_PRIME & 13 Nov. 2016 &  185 & -88.9 & 18.3\\
		CIRS\_248TI\_FIRNADCMP002\_PRIME & 14 Nov. 2016 &  186 & 30.3 & 17.4\\
		CIRS\_250TI\_FIRNADCMP002\_PRIME & 30 Nov. 2016 &  219 & -19.8 & 28.4\\			
		CIRS\_259TI\_COMPMAP001\_PIE & 01 Feb. 2017 &  302 & -69.0 & 20.6\\
		CIRS\_270TI\_FIRNADCMP001\_PRIME & 21 Apr. 2017 &  166 & -74.7 & 25.4\\			
		CIRS\_283TI\_COMPMAP001\_PRIME* & 10 Jul. 2017 &  114 & 60.0 & 26.5\\
		CIRS\_283TI\_COMPMAP001\_PRIME* & 10 Jul. 2017 &  134 & 67.5 & 24.7\\
		CIRS\_287TI\_COMPMAP001\_PIE & 11 Aug. 2017 &  305 & 88.9 & 9.3\\		
		CIRS\_288TI\_COMPMAP002\_PIE & 11 Aug. 2017 &  269 & 66.7 & 23.7\\
		CIRS\_292TI\_COMPMAP001\_PRIME & 12 Sep. 2017 &  192 & 70.4 & 19.2\\

	\hline
\end{longtable}

\twocolumn
\section*{References}	
\bibliographystyle{elsarticle-harv} 
\bibliography{biblio}

\begin{thebibliography}{44}
\expandafter\ifx\csname natexlab\endcsname\relax\def\natexlab#1{#1}\fi
\expandafter\ifx\csname url\endcsname\relax
  \def\url#1{\texttt{#1}}\fi
\expandafter\ifx\csname urlprefix\endcsname\relax\def\urlprefix{URL }\fi

\bibitem[{{Achterberg} et~al.(2008){Achterberg}, {Conrath}, {Gierasch},
  {Flasar}, and {Nixon}}]{Achterberg2008}
{Achterberg}, R.~K., {Conrath}, B.~J., {Gierasch}, P.~J., {Flasar}, F.~M.,
  {Nixon}, C.~A., Mar. 2008. {Titan's middle-atmospheric temperatures and
  dynamics observed by the Cassini Composite Infrared Spectrometer}. \icarus
  194, 263--277.

\bibitem[{{Achterberg} et~al.(2011){Achterberg}, {Gierasch}, {Conrath},
  {Michael Flasar}, and {Nixon}}]{Achterberg2011}
{Achterberg}, R.~K., {Gierasch}, P.~J., {Conrath}, B.~J., {Michael Flasar}, F.,
  {Nixon}, C.~A., Jan. 2011. {Temporal variations of Titan's middle-atmospheric
  temperatures from 2004 to 2009 observed by Cassini/CIRS}. \icarus 211,
  686--698.

\bibitem[{{Anderson} et~al.(2012){Anderson}, {Samuelson}, and
  {Achterberg}}]{Anderson2012}
{Anderson}, C., {Samuelson}, R., {Achterberg}, R., Apr. 2012. {Titan's
  stratospheric condensibles at high northern latitudes during northern
  winter}. In: {Cottini}, V., {Nixon}, C., {Lorenz}, R. (Eds.), Titan Through
  Time; Unlocking Titan's Past, Present and Future. p.~59.

\bibitem[{{Anderson} and {Samuelson}(2011)}]{Anderson2011}
{Anderson}, C.~M., {Samuelson}, R.~E., Apr. 2011. {Titan's aerosol and
  stratospheric ice opacities between 18 and 500 {$\mu$}m: Vertical and
  spectral characteristics from Cassini CIRS}. \icarus 212, 762--778.

\bibitem[{{Astropy Collaboration} et~al.(2013){Astropy Collaboration},
  {Robitaille}, {Tollerud}, {Greenfield}, {Droettboom}, {Bray}, {Aldcroft},
  {Davis}, {Ginsburg}, {Price-Whelan}, {Kerzendorf}, {Conley}, {Crighton},
  {Barbary}, {Muna}, {Ferguson}, {Grollier}, {Parikh}, {Nair}, {Unther},
  {Deil}, {Woillez}, {Conseil}, {Kramer}, {Turner}, {Singer}, {Fox}, {Weaver},
  {Zabalza}, {Edwards}, {Azalee Bostroem}, {Burke}, {Casey}, {Crawford},
  {Dencheva}, {Ely}, {Jenness}, {Labrie}, {Lim}, {Pierfederici}, {Pontzen},
  {Ptak}, {Refsdal}, {Servillat}, and {Streicher}}]{2013A&A...558A..33A}
{Astropy Collaboration}, {Robitaille}, T.~P., {Tollerud}, E.~J., {Greenfield},
  P., {Droettboom}, M., {Bray}, E., {Aldcroft}, T., {Davis}, M., {Ginsburg},
  A., {Price-Whelan}, A.~M., {Kerzendorf}, W.~E., {Conley}, A., {Crighton}, N.,
  {Barbary}, K., {Muna}, D., {Ferguson}, H., {Grollier}, F., {Parikh}, M.~M.,
  {Nair}, P.~H., {Unther}, H.~M., {Deil}, C., {Woillez}, J., {Conseil}, S.,
  {Kramer}, R., {Turner}, J.~E.~H., {Singer}, L., {Fox}, R., {Weaver}, B.~A.,
  {Zabalza}, V., {Edwards}, Z.~I., {Azalee Bostroem}, K., {Burke}, D.~J.,
  {Casey}, A.~R., {Crawford}, S.~M., {Dencheva}, N., {Ely}, J., {Jenness}, T.,
  {Labrie}, K., {Lim}, P.~L., {Pierfederici}, F., {Pontzen}, A., {Ptak}, A.,
  {Refsdal}, B., {Servillat}, M., {Streicher}, O., Oct. 2013. {Astropy: A
  community Python package for astronomy}. \aap 558, A33.

\bibitem[{{Bampasidis} et~al.(2012){Bampasidis}, {Coustenis}, {Achterberg},
  {Vinatier}, {Lavvas}, {Nixon}, {Jennings}, {Teanby}, {Flasar}, {Carlson},
  {Moussas}, {Preka-Papadema}, {Romani}, {Guandique}, and
  {Stamogiorgos}}]{Bampasidis2012}
{Bampasidis}, G., {Coustenis}, A., {Achterberg}, R.~K., {Vinatier}, S.,
  {Lavvas}, P., {Nixon}, C.~A., {Jennings}, D.~E., {Teanby}, N.~A., {Flasar},
  F.~M., {Carlson}, R.~C., {Moussas}, X., {Preka-Papadema}, P., {Romani},
  P.~N., {Guandique}, E.~A., {Stamogiorgos}, S., Dec. 2012. {Thermal and
  Chemical Structure Variations in Titan's Stratosphere during the Cassini
  Mission}. \apj 760, 144.

\bibitem[{{Barth}(2017)}]{Barth2017}
{Barth}, E.~L., Mar. 2017. {Modeling survey of ices in Titan's stratosphere}.
  \planss 137, 20--31.

\bibitem[{{B{\'e}zard} et~al.(2018){B{\'e}zard}, {Vinatier}, and
  {Achterberg}}]{Bezard2018}
{B{\'e}zard}, B., {Vinatier}, S., {Achterberg}, R.~K., Mar. 2018. {Seasonal
  radiative modeling of Titan's stratospheric temperatures at low latitudes}.
  \icarus 302, 437--450.

\bibitem[{{Cottini} et~al.(2012){Cottini}, {Nixon}, {Jennings}, {Anderson},
  {Gorius}, {Bjoraker}, {Coustenis}, {Teanby}, {Achterberg}, {B{\'e}zard}, {de
  Kok}, {Lellouch}, {Irwin}, {Flasar}, and {Bampasidis}}]{Cottini2012}
{Cottini}, V., {Nixon}, C.~A., {Jennings}, D.~E., {Anderson}, C.~M., {Gorius},
  N., {Bjoraker}, G.~L., {Coustenis}, A., {Teanby}, N.~A., {Achterberg}, R.~K.,
  {B{\'e}zard}, B., {de Kok}, R., {Lellouch}, E., {Irwin}, P.~G.~J., {Flasar},
  F.~M., {Bampasidis}, G., Aug. 2012. {Water vapor in Titan's stratosphere from
  Cassini CIRS far-infrared spectra}. \icarus 220, 855--862.

\bibitem[{{Coustenis} et~al.(2007){Coustenis}, {Achterberg}, {Conrath},
  {Jennings}, {Marten}, {Gautier}, {Nixon}, {Flasar}, {Teanby}, {B{\'e}zard},
  {Samuelson}, {Carlson}, {Lellouch}, {Bjoraker}, {Romani}, {Taylor}, {Irwin},
  {Fouchet}, {Hubert}, {Orton}, {Kunde}, {Vinatier}, {Mondellini}, {Abbas}, and
  {Courtin}}]{Coustenis2007}
{Coustenis}, A., {Achterberg}, R.~K., {Conrath}, B.~J., {Jennings}, D.~E.,
  {Marten}, A., {Gautier}, D., {Nixon}, C.~A., {Flasar}, F.~M., {Teanby},
  N.~A., {B{\'e}zard}, B., {Samuelson}, R.~E., {Carlson}, R.~C., {Lellouch},
  E., {Bjoraker}, G.~L., {Romani}, P.~N., {Taylor}, F.~W., {Irwin}, P.~G.~J.,
  {Fouchet}, T., {Hubert}, A., {Orton}, G.~S., {Kunde}, V.~G., {Vinatier}, S.,
  {Mondellini}, J., {Abbas}, M.~M., {Courtin}, R., Jul. 2007. {The composition
  of Titan's stratosphere from Cassini/CIRS mid-infrared spectra}. \icarus 189,
  35--62.

\bibitem[{{Coustenis} et~al.(2016){Coustenis}, {Jennings}, {Achterberg},
  {Bampasidis}, {Lavvas}, {Nixon}, {Teanby}, {Anderson}, {Cottini}, and
  {Flasar}}]{Coustenis2016}
{Coustenis}, A., {Jennings}, D.~E., {Achterberg}, R.~K., {Bampasidis}, G.,
  {Lavvas}, P., {Nixon}, C.~A., {Teanby}, N.~A., {Anderson}, C.~M., {Cottini},
  V., {Flasar}, F.~M., May 2016. {Titan's temporal evolution in stratospheric
  trace gases near the poles}. \icarus 270, 409--420.

\bibitem[{{Coustenis} et~al.(1999){Coustenis}, {Schmitt}, {Khanna}, and
  {Trotta}}]{Coustenis1999}
{Coustenis}, A., {Schmitt}, B., {Khanna}, R.~K., {Trotta}, F., Oct. 1999.
  {Plausible condensates in Titan's stratosphere from Voyager infrared
  spectra}. \planss 47, 1305--1329.

\bibitem[{{de Kok} et~al.(2007){de Kok}, {Irwin}, {Teanby}, {Nixon},
  {Jennings}, {Fletcher}, {Howett}, {Calcutt}, {Bowles}, {Flasar}, and
  {Taylor}}]{deKok2007}
{de Kok}, R., {Irwin}, P.~G.~J., {Teanby}, N.~A., {Nixon}, C.~A., {Jennings},
  D.~E., {Fletcher}, L., {Howett}, C., {Calcutt}, S.~B., {Bowles}, N.~E.,
  {Flasar}, F.~M., {Taylor}, F.~W., Nov. 2007. {Characteristics of Titan's
  stratospheric aerosols and condensate clouds from Cassini CIRS far-infrared
  spectra}. \icarus 191, 223--235.

\bibitem[{{de Kok} et~al.(2010){de Kok}, {Irwin}, {Teanby}, {Vinatier}, {Tosi},
  {Negr{\~a}o}, {Osprey}, {Adriani}, {Moriconi}, and {Coradini}}]{deKok2010b}
{de Kok}, R., {Irwin}, P.~G.~J., {Teanby}, N.~A., {Vinatier}, S., {Tosi}, F.,
  {Negr{\~a}o}, A., {Osprey}, S., {Adriani}, A., {Moriconi}, M.~L., {Coradini},
  A., May 2010. {A tropical haze band in Titan's stratosphere}. \icarus 207,
  485--490.

\bibitem[{{de Kok} et~al.(2014){de Kok}, {Teanby}, {Maltagliati}, {Irwin}, and
  {Vinatier}}]{deKok2014}
{de Kok}, R.~J., {Teanby}, N.~A., {Maltagliati}, L., {Irwin}, P. G.~J.,
  {Vinatier}, S., Oct. 2014. {HCN ice in Titan's high-altitude southern polar
  cloud}. \nat 514, 65--67.

\bibitem[{{Flasar} et~al.(2004){Flasar}, {Kunde}, {Abbas}, {Achterberg}, {Ade},
  {Barucci}, {B{\'e}zard}, {Bjoraker}, {Brasunas}, {Calcutt}, {Carlson},
  {C{\'e}sarsky}, {Conrath}, {Coradini}, {Courtin}, {Coustenis}, {Edberg},
  {Edgington}, {Ferrari}, {Fouchet}, {Gautier}, {Gierasch}, {Grossman},
  {Irwin}, {Jennings}, {Lellouch}, {Mamoutkine}, {Marten}, {Meyer}, {Nixon},
  {Orton}, {Owen}, {Pearl}, {Prang{\'e}}, {Raulin}, {Read}, {Romani},
  {Samuelson}, {Segura}, {Showalter}, {Simon-Miller}, {Smith}, {Spencer},
  {Spilker}, and {Taylor}}]{Flasar2004}
{Flasar}, F.~M., {Kunde}, V.~G., {Abbas}, M.~M., {Achterberg}, R.~K., {Ade},
  P., {Barucci}, A., {B{\'e}zard}, B., {Bjoraker}, G.~L., {Brasunas}, J.~C.,
  {Calcutt}, S., {Carlson}, R., {C{\'e}sarsky}, C.~J., {Conrath}, B.~J.,
  {Coradini}, A., {Courtin}, R., {Coustenis}, A., {Edberg}, S., {Edgington},
  S., {Ferrari}, C., {Fouchet}, T., {Gautier}, D., {Gierasch}, P.~J.,
  {Grossman}, K., {Irwin}, P., {Jennings}, D.~E., {Lellouch}, E., {Mamoutkine},
  A.~A., {Marten}, A., {Meyer}, J.~P., {Nixon}, C.~A., {Orton}, G.~S., {Owen},
  T.~C., {Pearl}, J.~C., {Prang{\'e}}, R., {Raulin}, F., {Read}, P.~L.,
  {Romani}, P.~N., {Samuelson}, R.~E., {Segura}, M.~E., {Showalter}, M.~R.,
  {Simon-Miller}, A.~A., {Smith}, M.~D., {Spencer}, J.~R., {Spilker}, L.~J.,
  {Taylor}, F.~W., Dec. 2004. {Exploring The Saturn System In The Thermal
  Infrared: The Composite Infrared Spectrometer}. \ssr 115, 169--297.

\bibitem[{{Fulchignoni} et~al.(2005){Fulchignoni}, {Ferri}, {Angrilli}, {Ball},
  {Bar-Nun}, {Barucci}, {Bettanini}, {Bianchini}, {Borucki}, {Colombatti},
  {Coradini}, {Coustenis}, {Debei}, {Falkner}, {Fanti}, {Flamini}, {Gaborit},
  {Grard}, {Hamelin}, {Harri}, {Hathi}, {Jernej}, {Leese}, {Lehto}, {Lion
  Stoppato}, {L{\'o}pez-Moreno}, {M{\"a}kinen}, {McDonnell}, {McKay},
  {Molina-Cuberos}, {Neubauer}, {Pirronello}, {Rodrigo}, {Saggin},
  {Schwingenschuh}, {Seiff}, {Sim{\~o}es}, {Svedhem}, {Tokano}, {Towner},
  {Trautner}, {Withers}, and {Zarnecki}}]{Fulchignoni2005}
{Fulchignoni}, M., {Ferri}, F., {Angrilli}, F., {Ball}, A.~J., {Bar-Nun}, A.,
  {Barucci}, M.~A., {Bettanini}, C., {Bianchini}, G., {Borucki}, W.,
  {Colombatti}, G., {Coradini}, M., {Coustenis}, A., {Debei}, S., {Falkner},
  P., {Fanti}, G., {Flamini}, E., {Gaborit}, V., {Grard}, R., {Hamelin}, M.,
  {Harri}, A.~M., {Hathi}, B., {Jernej}, I., {Leese}, M.~R., {Lehto}, A., {Lion
  Stoppato}, P.~F., {L{\'o}pez-Moreno}, J.~J., {M{\"a}kinen}, T., {McDonnell},
  J.~A.~M., {McKay}, C.~P., {Molina-Cuberos}, G., {Neubauer}, F.~M.,
  {Pirronello}, V., {Rodrigo}, R., {Saggin}, B., {Schwingenschuh}, K., {Seiff},
  A., {Sim{\~o}es}, F., {Svedhem}, H., {Tokano}, T., {Towner}, M.~C.,
  {Trautner}, R., {Withers}, P., {Zarnecki}, J.~C., Dec. 2005. {In situ
  measurements of the physical characteristics of Titan's environment}. \nat
  438, 785--791.

\bibitem[{Hunter(2007)}]{Hunter:2007}
Hunter, J.~D., 2007. Matplotlib: A 2d graphics environment. Computing In
  Science \& Engineering 9~(3), 90--95.

\bibitem[{{Irwin} et~al.(2008){Irwin}, {Teanby}, {de Kok}, {Fletcher},
  {Howett}, {Tsang}, {Wilson}, {Calcutt}, {Nixon}, and {Parrish}}]{Irwin2008}
{Irwin}, P.~G.~J., {Teanby}, N.~A., {de Kok}, R., {Fletcher}, L.~N., {Howett},
  C.~J.~A., {Tsang}, C.~C.~C., {Wilson}, C.~F., {Calcutt}, S.~B., {Nixon},
  C.~A., {Parrish}, P.~D., Apr. 2008. {The NEMESIS planetary atmosphere
  radiative transfer and retrieval tool}. \jqsrt 109, 1136--1150.

\bibitem[{{Jennings} et~al.(2015){Jennings}, {Achterberg}, {Cottini},
  {Anderson}, {Flasar}, {Nixon}, {Bjoraker}, {Kunde}, {Carlson}, {Guandique},
  {Kaelberer}, {Tingley}, {Albright}, {Segura}, {de Kok}, {Coustenis},
  {Vinatier}, {Bampasidis}, {Teanby}, and {Calcutt}}]{Jennings2015}
{Jennings}, D.~E., {Achterberg}, R.~K., {Cottini}, V., {Anderson}, C.~M.,
  {Flasar}, F.~M., {Nixon}, C.~A., {Bjoraker}, G.~L., {Kunde}, V.~G.,
  {Carlson}, R.~C., {Guandique}, E., {Kaelberer}, M.~S., {Tingley}, J.~S.,
  {Albright}, S.~A., {Segura}, M.~E., {de Kok}, R., {Coustenis}, A.,
  {Vinatier}, S., {Bampasidis}, G., {Teanby}, N.~A., {Calcutt}, S., May 2015.
  {Evolution of the Far-infrared Cloud at Titan's South Pole}. \apjl 804, L34.

\bibitem[{{Jennings} et~al.(2012){Jennings}, {Anderson}, {Samuelson}, {Flasar},
  {Nixon}, {Kunde}, {Achterberg}, {Cottini}, {de Kok}, {Coustenis}, {Vinatier},
  and {Calcutt}}]{Jennings2012}
{Jennings}, D.~E., {Anderson}, C.~M., {Samuelson}, R.~E., {Flasar}, F.~M.,
  {Nixon}, C.~A., {Kunde}, V.~G., {Achterberg}, R.~K., {Cottini}, V., {de Kok},
  R., {Coustenis}, A., {Vinatier}, S., {Calcutt}, S.~B., Jul. 2012. {Seasonal
  Disappearance of Far-infrared Haze in Titan's Stratosphere}. \apjl 754, L3.

\bibitem[{{Lebonnois} et~al.(2012){Lebonnois}, {Burgalat}, {Rannou}, and
  {Charnay}}]{Lebonnois2012a}
{Lebonnois}, S., {Burgalat}, J., {Rannou}, P., {Charnay}, B., Mar. 2012. {Titan
  global climate model: A new 3-dimensional version of the IPSL Titan GCM}.
  \icarus 218, 707--722.

\bibitem[{{Lellouch} et~al.(2014){Lellouch}, {B{\'e}zard}, {Flasar},
  {Vinatier}, {Achterberg}, {Nixon}, {Bjoraker}, and {Gorius}}]{Lellouch2014}
{Lellouch}, E., {B{\'e}zard}, B., {Flasar}, F.~M., {Vinatier}, S.,
  {Achterberg}, R., {Nixon}, C.~A., {Bjoraker}, G.~L., {Gorius}, N., Mar. 2014.
  {The distribution of methane in Titan's stratosphere from Cassini/CIRS
  observations}. \icarus 231, 323--337.

\bibitem[{{Lora} et~al.(2015){Lora}, {Lunine}, and {Russell}}]{Lora2015}
{Lora}, J.~M., {Lunine}, J.~I., {Russell}, J.~L., Apr. 2015. {GCM simulations
  of Titan's middle and lower atmosphere and comparison to observations}.
  \icarus 250, 516--528.

\bibitem[{{Maltagliati} et~al.(2015){Maltagliati}, {B{\'e}zard}, {Vinatier},
  {Hedman}, {Lellouch}, {Nicholson}, {Sotin}, {de Kok}, and
  {Sicardy}}]{Maltagliati2015}
{Maltagliati}, L., {B{\'e}zard}, B., {Vinatier}, S., {Hedman}, M.~M.,
  {Lellouch}, E., {Nicholson}, P.~D., {Sotin}, C., {de Kok}, R.~J., {Sicardy},
  B., Mar. 2015. {Titan's atmosphere as observed by Cassini/VIMS solar
  occultations: CH$_{4}$, CO and evidence for C$_{2}$H$_{6}$ absorption}.
  \icarus 248, 1--24.

\bibitem[{{Molter} et~al.(2016){Molter}, {Nixon}, {Cordiner}, {Serigano},
  {Irwin}, {Teanby}, {Charnley}, and {Lindberg}}]{Molter2016}
{Molter}, E.~M., {Nixon}, C.~A., {Cordiner}, M.~A., {Serigano}, J., {Irwin},
  P.~G.~J., {Teanby}, N.~A., {Charnley}, S.~B., {Lindberg}, J.~E., Aug. 2016.
  {ALMA Observations of HCN and Its Isotopologues on Titan}. \aj 152, 42.

\bibitem[{{Newman} et~al.(2011){Newman}, {Lee}, {Lian}, {Richardson}, and
  {Toigo}}]{Newman2011}
{Newman}, C.~E., {Lee}, C., {Lian}, Y., {Richardson}, M.~I., {Toigo}, A.~D.,
  Jun. 2011. {Stratospheric superrotation in the TitanWRF model}. \icarus 213,
  636--654.

\bibitem[{{Niemann} et~al.(2010){Niemann}, {Atreya}, {Demick}, {Gautier},
  {Haberman}, {Harpold}, {Kasprzak}, {Lunine}, {Owen}, and
  {Raulin}}]{Niemann2010}
{Niemann}, H.~B., {Atreya}, S.~K., {Demick}, J.~E., {Gautier}, D., {Haberman},
  J.~A., {Harpold}, D.~N., {Kasprzak}, W.~T., {Lunine}, J.~I., {Owen}, T.~C.,
  {Raulin}, F., Dec. 2010. {Composition of Titan's lower atmosphere and simple
  surface volatiles as measured by the Cassini-Huygens probe gas chromatograph
  mass spectrometer experiment}. Journal of Geophysical Research (Planets) 115,
  E12006.

\bibitem[{{Nixon} et~al.(2012){Nixon}, {Temelso}, {Vinatier}, {Teanby},
  {B{\'e}zard}, {Achterberg}, {Mandt}, {Sherrill}, {Irwin}, {Jennings},
  {Romani}, {Coustenis}, and {Flasar}}]{Nixon2012}
{Nixon}, C.~A., {Temelso}, B., {Vinatier}, S., {Teanby}, N.~A., {B{\'e}zard},
  B., {Achterberg}, R.~K., {Mandt}, K.~E., {Sherrill}, C.~D., {Irwin},
  P.~G.~J., {Jennings}, D.~E., {Romani}, P.~N., {Coustenis}, A., {Flasar},
  F.~M., Apr. 2012. {Isotopic Ratios in Titan's Methane: Measurements and
  Modeling}. \apj 749, 159.

\bibitem[{{Rothman} et~al.(2013){Rothman}, {Gordon}, {Babikov}, {Barbe}, {Chris
  Benner}, {Bernath}, {Birk}, {Bizzocchi}, {Boudon}, {Brown}, {Campargue},
  {Chance}, {Cohen}, {Coudert}, {Devi}, {Drouin}, {Fayt}, {Flaud}, {Gamache},
  {Harrison}, {Hartmann}, {Hill}, {Hodges}, {Jacquemart}, {Jolly}, {Lamouroux},
  {Le Roy}, {Li}, {Long}, {Lyulin}, {Mackie}, {Massie}, {Mikhailenko},
  {M{\"u}ller}, {Naumenko}, {Nikitin}, {Orphal}, {Perevalov}, {Perrin},
  {Polovtseva}, {Richard}, {Smith}, {Starikova}, {Sung}, {Tashkun}, {Tennyson},
  {Toon}, {Tyuterev}, and {Wagner}}]{Rothman2013}
{Rothman}, L.~S., {Gordon}, I.~E., {Babikov}, Y., {Barbe}, A., {Chris Benner},
  D., {Bernath}, P.~F., {Birk}, M., {Bizzocchi}, L., {Boudon}, V., {Brown},
  L.~R., {Campargue}, A., {Chance}, K., {Cohen}, E.~A., {Coudert}, L.~H.,
  {Devi}, V.~M., {Drouin}, B.~J., {Fayt}, A., {Flaud}, J.-M., {Gamache}, R.~R.,
  {Harrison}, J.~J., {Hartmann}, J.-M., {Hill}, C., {Hodges}, J.~T.,
  {Jacquemart}, D., {Jolly}, A., {Lamouroux}, J., {Le Roy}, R.~J., {Li}, G.,
  {Long}, D.~A., {Lyulin}, O.~M., {Mackie}, C.~J., {Massie}, S.~T.,
  {Mikhailenko}, S., {M{\"u}ller}, H.~S.~P., {Naumenko}, O.~V., {Nikitin},
  A.~V., {Orphal}, J., {Perevalov}, V., {Perrin}, A., {Polovtseva}, E.~R.,
  {Richard}, C., {Smith}, M.~A.~H., {Starikova}, E., {Sung}, K., {Tashkun}, S.,
  {Tennyson}, J., {Toon}, G.~C., {Tyuterev}, V.~G., {Wagner}, G., Nov. 2013.
  {The HITRAN2012 molecular spectroscopic database}. \jqsrt 130, 4--50.

\bibitem[{{Schinder} et~al.(2011){Schinder}, {Flasar}, {Marouf}, {French},
  {McGhee}, {Kliore}, {Rappaport}, {Barbinis}, {Fleischman}, and
  {Anabtawi}}]{SchinderFlasarMaroufEtAl2011}
{Schinder}, P.~J., {Flasar}, F.~M., {Marouf}, E.~A., {French}, R.~G., {McGhee},
  C.~A., {Kliore}, A.~J., {Rappaport}, N.~J., {Barbinis}, E., {Fleischman}, D.,
  {Anabtawi}, A., Oct. 2011. {The structure of Titan's atmosphere from Cassini
  radio occultations}. \icarus 215, 460--474.

\bibitem[{{Schinder} et~al.(2012){Schinder}, {Flasar}, {Marouf}, {French},
  {McGhee}, {Kliore}, {Rappaport}, {Barbinis}, {Fleischman}, and
  {Anabtawi}}]{Schinder2012}
{Schinder}, P.~J., {Flasar}, F.~M., {Marouf}, E.~A., {French}, R.~G., {McGhee},
  C.~A., {Kliore}, A.~J., {Rappaport}, N.~J., {Barbinis}, E., {Fleischman}, D.,
  {Anabtawi}, A., Nov. 2012. {The structure of Titan's atmosphere from Cassini
  radio occultations: Occultations from the Prime and Equinox missions}.
  \icarus 221, 1020--1031.

\bibitem[{{Strobel} et~al.(2010){Strobel}, {Atreya}, {B{\'e}zard}, {Ferri},
  {Flasar}, {Fulchignoni}, {Lellouch}, and {M{\"u}ller-Wodarg}}]{Strobel2010}
{Strobel}, D.~F., {Atreya}, S.~K., {B{\'e}zard}, B., {Ferri}, F., {Flasar},
  F.~M., {Fulchignoni}, M., {Lellouch}, E., {M{\"u}ller-Wodarg}, I., 2010.
  {Atmospheric Structure and Composition}. p. 235.

\bibitem[{{Sylvestre} et~al.(2018){Sylvestre}, {Teanby}, {Vinatier},
  {Lebonnois}, and {Irwin}}]{Sylvestre2018}
{Sylvestre}, M., {Teanby}, N.~A., {Vinatier}, S., {Lebonnois}, S., {Irwin},
  P.~G.~J., Jan. 2018. {Seasonal evolution of C$_{2}$N$_{2}$, C$_{3}$H$_{4}$,
  and C$_{4}$H$_{2}$ abundances in Titan's lower stratosphere}. \aap 609, A64.

\bibitem[{{Teanby} et~al.(2017){Teanby}, {B{\'e}zard}, {Vinatier}, {Sylvestre},
  {Nixon}, {Irwin}, {de Kok}, {Calcutt}, and {Flasar}}]{Teanby2017}
{Teanby}, N.~A., {B{\'e}zard}, B., {Vinatier}, S., {Sylvestre}, M., {Nixon},
  C.~A., {Irwin}, P.~G.~J., {de Kok}, R.~J., {Calcutt}, S.~B., {Flasar}, F.~M.,
  Nov. 2017. {The formation and evolution of Titan's winter polar vortex}.
  Nature Communications 8, 1586.

\bibitem[{{Teanby} et~al.(2009){Teanby}, {Irwin}, {de Kok}, {Jolly},
  {B{\'e}zard}, {Nixon}, and {Calcutt}}]{Teanby2009}
{Teanby}, N.~A., {Irwin}, P.~G.~J., {de Kok}, R., {Jolly}, A., {B{\'e}zard},
  B., {Nixon}, C.~A., {Calcutt}, S.~B., Aug. 2009. {Titan's stratospheric C
  $_{2}$N $_{2}$, C $_{3}$H $_{4}$, and C $_{4}$H $_{2}$ abundances from
  Cassini/CIRS far-infrared spectra}. \icarus 202, 620--631.

\bibitem[{{Teanby} et~al.(2007){Teanby}, {Irwin}, {de Kok}, {Vinatier},
  {B{\'e}zard}, {Nixon}, {Flasar}, {Calcutt}, {Bowles}, {Fletcher}, {Howett},
  and {Taylor}}]{Teanby2007b}
{Teanby}, N.~A., {Irwin}, P.~G.~J., {de Kok}, R., {Vinatier}, S., {B{\'e}zard},
  B., {Nixon}, C.~A., {Flasar}, F.~M., {Calcutt}, S.~B., {Bowles}, N.~E.,
  {Fletcher}, L., {Howett}, C., {Taylor}, F.~W., Feb. 2007. {Vertical profiles
  of HCN, HC <SUB>3</SUB>N, and C <SUB>2</SUB>H <SUB>2</SUB> in Titan's
  atmosphere derived from Cassini/CIRS data}. \icarus 186, 364--384.

\bibitem[{{Teanby} et~al.(2012){Teanby}, {Irwin}, {Nixon}, {de Kok},
  {Vinatier}, {Coustenis}, {Sefton-Nash}, {Calcutt}, and {Flasar}}]{Teanby2012}
{Teanby}, N.~A., {Irwin}, P.~G.~J., {Nixon}, C.~A., {de Kok}, R., {Vinatier},
  S., {Coustenis}, A., {Sefton-Nash}, E., {Calcutt}, S.~B., {Flasar}, F.~M.,
  Nov. 2012. {Active upper-atmosphere chemistry and dynamics from polar
  circulation reversal on Titan}. \nat 491, 732--735.

\bibitem[{{Tomasko} et~al.(2008){Tomasko}, {Doose}, {Engel}, {Dafoe}, {West},
  {Lemmon}, {Karkoschka}, and {See}}]{Tomasko2008b}
{Tomasko}, M.~G., {Doose}, L., {Engel}, S., {Dafoe}, L.~E., {West}, R.,
  {Lemmon}, M., {Karkoschka}, E., {See}, C., Apr. 2008. {A model of Titan's
  aerosols based on measurements made inside the atmosphere}. \planss 56,
  669--707.

\bibitem[{{Vatant d'Ollone} et~al.(2017){Vatant d'Ollone}, {Lebonnois}, and
  {Guerlet}}]{Vatantd'Ollone2017}
{Vatant d'Ollone}, J., {Lebonnois}, S., {Guerlet}, S., Apr. 2017. {Modelling of
  Titan's middle atmosphere with the IPSL climate model}. In: EGU General
  Assembly Conference Abstracts. Vol.~19. p. 10169.

\bibitem[{{Vinatier} et~al.(2007){Vinatier}, {B{\'e}zard}, {Fouchet}, {Teanby},
  {de Kok}, {Irwin}, {Conrath}, {Nixon}, {Romani}, {Flasar}, and
  {Coustenis}}]{Vinatier2007}
{Vinatier}, S., {B{\'e}zard}, B., {Fouchet}, T., {Teanby}, N.~A., {de Kok}, R.,
  {Irwin}, P. G.~J., {Conrath}, B.~J., {Nixon}, C.~A., {Romani}, P.~N.,
  {Flasar}, F.~M., {Coustenis}, A., May 2007. {Vertical abundance profiles of
  hydrocarbons in Titan's atmosphere at 15$^{\circ}$S and 80$^{\circ}$ N
  retrieved from Cassini/CIRS spectra}. \icarus 188, 120--138.

\bibitem[{{Vinatier} et~al.(2015){Vinatier}, {B{\'e}zard}, {Lebonnois},
  {Teanby}, {Achterberg}, {Gorius}, {Mamoutkine}, {Guandique}, {Jolly},
  {Jennings}, and {Flasar}}]{Vinatier2015}
{Vinatier}, S., {B{\'e}zard}, B., {Lebonnois}, S., {Teanby}, N.~A.,
  {Achterberg}, R.~K., {Gorius}, N., {Mamoutkine}, A., {Guandique}, E.,
  {Jolly}, A., {Jennings}, D.~E., {Flasar}, F.~M., Apr. 2015. {Seasonal
  variations in Titan's middle atmosphere during the northern spring derived
  from Cassini/CIRS observations}. \icarus 250, 95--115.

\bibitem[{{Vinatier} et~al.(2012){Vinatier}, {Rannou}, {Anderson},
  {B{\'e}zard}, {de Kok}, and {Samuelson}}]{Vinatier2012}
{Vinatier}, S., {Rannou}, P., {Anderson}, C.~M., {B{\'e}zard}, B., {de Kok},
  R., {Samuelson}, R.~E., May 2012. {Optical constants of Titan's stratospheric
  aerosols in the 70-1500 cm$^{-1}$ spectral range constrained by Cassini/CIRS
  observations}. \icarus 219, 5--12.

\bibitem[{{Vinatier} et~al.(2018){Vinatier}, {Schmitt}, {B{\'e}zard}, {Rannou},
  {Dauphin}, {de Kok}, {Jennings}, and {Flasar}}]{Vinatier2018}
{Vinatier}, S., {Schmitt}, B., {B{\'e}zard}, B., {Rannou}, P., {Dauphin}, C.,
  {de Kok}, R., {Jennings}, D.~E., {Flasar}, F.~M., Aug. 2018. {Study of
  Titan's fall southern stratospheric polar cloud composition with
  Cassini/CIRS: Detection of benzene ice}. \icarus 310, 89--104.

\end{thebibliography}

\end{document}